\documentclass[twocolumn]{aastex631} 

\providecommand{\bjdtdb}{\ensuremath{\rm {BJD_{TDB}}}}

\providecommand{\msun}{\ensuremath{M_\odot}}
\providecommand{\rsun}{\ensuremath{R_\odot}}
\providecommand{\lsun}{\ensuremath{L_\odot}}
\providecommand{\msun}{\ensuremath{M}}
\providecommand{\rsun}{\ensuremath{R}}
\providecommand{\lsun}{\ensuremath{L}}

\providecommand{\fave}{\langle F \rangle} 
\providecommand{\fluxcgs}{10$^9$ erg s$^{-1}$ cm$^{-2}$}

\providecommand{\arcsec}{$^{\prime \prime}$}
\providecommand{\arcmin}{$^{\prime}$}

\usepackage{makecell}

\begin{document} 



\title{TOI-6038~A~b: A dense sub-Saturn in the transition regime between the Neptunian ridge and savanna} 
\author[0000-0001-8998-3223]{Sanjay Baliwal}
\affiliation{Astronomy $\&$ Astrophysics Division, Physical Research Laboratory, Ahmedabad 380009, India}
\affiliation{Indian Institute of Technology, Gandhinagar 382355, India}

\author[0000-0001-8983-5300]{Rishikesh Sharma}
\affiliation{Astronomy $\&$ Astrophysics Division, Physical Research Laboratory, Ahmedabad 380009, India}

\author[0000-0002-3815-8407]{Abhijit Chakraborty}
\affiliation{Astronomy $\&$ Astrophysics Division, Physical Research Laboratory, Ahmedabad 380009, India}

\author[0009-0000-4834-5612]{K. J. Nikitha}
\affiliation{Astronomy $\&$ Astrophysics Division, Physical Research Laboratory, Ahmedabad 380009, India}

\author[0000-0001-7439-3618]{A. Castro-González}
\affiliation{Centro de Astrobiología, CSIC-INTA, Camino Bajo del Castillo s/n, 28692 Villanueva de la Cañada, Madrid, Spain}

\author[0000-0002-5181-0463]{Hareesh G. Bhaskar}
\affiliation{Technion - Israel Institute of Technology, Department of Physics, Haifa 3200003 Israel}

\author[0000-0003-0335-6435]{Akanksha Khandelwal}
\affiliation{Astronomy $\&$ Astrophysics Division, Physical Research Laboratory, Ahmedabad 380009, India}
\affiliation{Instituto de Astronomía, Universidad Nacional Autónoma de México, Av. Universidad No. 3000, Ciudad de México, 04510, Mexico}

\author[0000-0001-9911-7388]{David W. Latham}
\affiliation{Center for Astrophysics | Harvard $\&$ Smithsonian, 60 Garden St., Cambridge MA 02138, USA}

\author[0000-0001-6637-5401]{Allyson Bieryla}
\affiliation{Center for Astrophysics | Harvard $\&$ Smithsonian, 60 Garden St., Cambridge MA 02138, USA}

\author[0000-0002-9148-034X]{Vincent Bourrier}
\affiliation{Observatoire Astronomique de l'Universit\'e de Gen\`eve, Chemin Pegasi 51b, CH-1290 Versoix, Switzerland}

\author[0000-0003-0670-5821]{Neelam J.S.S.V. Prasad}
\affiliation{Astronomy $\&$ Astrophysics Division, Physical Research Laboratory, Ahmedabad 380009, India}

\author[0000-0003-1373-4583]{Kapil K. Bharadwaj}
\affiliation{Astronomy $\&$ Astrophysics Division, Physical Research Laboratory, Ahmedabad 380009, India}

\author{Kevikumar A. Lad}
\affiliation{Astronomy $\&$ Astrophysics Division, Physical Research Laboratory, Ahmedabad 380009, India}

\author[0009-0001-3782-4308]{Ashirbad Nayak}
\affiliation{Astronomy $\&$ Astrophysics Division, Physical Research Laboratory, Ahmedabad 380009, India}

\author[0000-0002-1457-4027]{Vishal Joshi}
\affiliation{Astronomy $\&$ Astrophysics Division, Physical Research Laboratory, Ahmedabad 380009, India}

\author[0000-0003-3773-5142]{Jason D. Eastman}
\affiliation{Center for Astrophysics | Harvard $\&$ Smithsonian, 60 Garden St., Cambridge MA 02138, USA}

\correspondingauthor{Sanjay Baliwal}
\email{sanjaybaliwal1998@gmail.com}


\begin{abstract}

We present the discovery and characterization of a sub-Saturn exoplanet, TOI-6038~A~b, using the PARAS-2 spectrograph. The planet orbits a bright ($m_V=9.9$), metal-rich late F-type star, TOI-6038~A, with $T_{\rm{eff}}=6110\pm100~\mathrm{K}$, $\log{g}=4.118^{+0.015}_{-0.025}$, and $[{\rm{Fe/H}}]=0.124^{+0.079}_{-0.077}$ dex. The system also contains a wide-orbit binary companion, TOI-6038~B, an early K-type star at a projected separation of $\approx3217$ AU. We combined radial velocity data from PARAS-2 with photometric data from the Transiting Exoplanet Survey Satellite (TESS) for joint modeling. TOI-6038~A~b has a mass of $78.5^{+9.5}_{-9.9}~M_\oplus$ and a radius of $6.41^{+0.20}_{-0.16}~R_\oplus$, orbiting in a circular orbit with a period of $5.8267311^{+0.0000074}_{-0.0000068}$ days. Internal structure modeling suggests that $\approx74\%$ of the planet's mass is composed of dense materials, such as rock and iron, forming a core, while the remaining mass consists of a low-density H/He envelope. TOI-6038~A~b lies at the transition regime between the recently identified Neptunian ridge and savanna. Having a density of $\rho_{\rm{P}}=1.62^{+0.23}_{-0.24}\rm~g\,cm^{-3}$, TOI-6038~A~b is compatible with the population of dense ridge planets ($\rho_{\rm{P}}\simeq$ 1.5-2.0 $\rm~g\,cm^{-3}$), which have been proposed to have reached their close-in locations through high-eccentricity tidal migration (HEM). First-order estimates suggest that the secular perturbations induced by TOI-6038~B may be insufficient to drive the HEM of TOI-6038~A~b. Therefore, it is not clear whether HEM driven by a still undetected companion, or early disk-driven migration, brought TOI-6038~A~b to its present-day close-in orbit. Interestingly, its bright host star makes TOI-6038~A~b a prime target for atmospheric escape and orbital architecture observations, which will help us to better understand its overall evolution.

\end{abstract}



\section{Introduction} \label{sec:intro}

The discovery of extra-solar planets has revealed numerous planetary populations with no analogs in the Solar System. Among these, close-in sub-Saturns  \citep[$P_{\rm orb}$ $<$ 30 days; $R_{\rm p}$ = 4–8 $R_\oplus$;][]{Petigura_2016} represent a particularly intriguing class. These planets have significant gaseous envelopes and are thought to form through core accretion similarly to the gas giant population, beyond the ice line, where planetary cores can reach a critical mass of approximately $10 \ M_\oplus$ capable of triggering runaway gas accretion \citep[e.g.][]{POLLACK_1996, Lee_2015}. However, their formation process must have been interrupted, probably due to late formation timing or an early disk dissipation \citep[e.g.][]{Mordasini_11, Mordasini_15}, resulting in final sizes intermediate between Neptune and Saturn. Sub-Saturns are thus thought to have migrated inwards to their present-day close-in orbits. Two main migration mechanisms have been proposed: high-eccentricity tidal migration \citep[HEM; e.g.][]{wuPlanetMigrationBinary2003,2011CeMDA.111..105C} and disk-driven migration \citep[e.g.][]{1979ApJ...233..857G,1996Natur.380..606L}. The presence of wide binary stellar companions or other planetary companions in some of these systems provides a plausible mechanism for initiating HEM through secular perturbations \citep{wuPlanetMigrationBinary2003,Ford_2008}. However, the efficiency of these perturbations in driving giant planets to very close orbits and the subsequent circularization process are not yet well understood. After migrating inwards or during the migration, these planets are thought to undergo atmospheric mass loss through photoevaporation or tidal heating mechanisms. It has been argued that due to high core mass in dense sub-Saturns, these planets may not favor the photoevaporation but would lose their mass through tidal heating, while the less dense sub-Saturns favor the photoevaporation hypothesis. In all, both atmospheric and dynamical processes are thought to play a critical role in shaping the population of close-in sub-Saturn exoplanets \citep[e.g.][]{2016ApJ...820L...8M,2017AJ....154..192G,2018Natur.553..477B,Owen2018,2022AJ....164..234V}.

A recent study by \cite{CastroGonzalez2024a} has unveiled a complex feature in the orbital period distribution of these planets, which is marked by distinct regions. They identified an over-density of super-Neptunes and sub-Saturns in the $\simeq$ 3.2 to 5.7 days orbital period range, resembling the well-known hot Jupiter pileup of larger planets \citep{1999ApJ...526..890C,2003A&A...407..369U,2005ApJ...623..472G,2006ApJ...646..505B}. This over-density was termed the Neptunian ridge since it separates the Neptunian desert \citep[i.e. a dearth of super-Neptunes and sub-Saturns in the shortest orbital periods;][]{2011A&A...528A...2B,2011ApJ...727L..44S,Lundkvist_16,Mazeh2016} and savanna \citep[i.e. a moderately populated region at larger orbital distances;][]{Bourrier2023}. \citet{CastroGonzalez2024b} found a significant dichotomy between the densities of the Neptunes in the ridge and savanna. Planets in the savanna have low densities ($\rm \sim 0.5 \, g\,cm^{-3}$), very rarely surpassing $\rm 1 g\,cm^{-3}$, while planets in the ridge frequently show densities as high as $\rm 1.5-2.0 \, \rm g\,cm^{-3}$, which cannot be explained through observational biases since denser planets are easier to detect. The very recent detection and characterization of the ridge planet TOI-6651~b \citep{toi6651} as one of the densest sub-Saturns known to date ($\rho_{\rm P}$ = $2.52^{+0.52}_{-0.44} \, \rm g\,cm^{-3}$) further supports this trend. The contrasting densities in the ridge and savanna are thought to be related to the formation and/or migration mechanisms leading to these two populations, but the predominant process has not been yet identified. It has also been observed that a large fraction of the planets in the ridge have misaligned and eccentric orbits, supporting HEM for their origins \citep{Correia_20, Bourrier2023}. 

The current number of sub-Saturns with precise measurements of mass and radius is limited. The various regimes of the exo-Neptunian landscape need to be further populated for a better understanding of their formation and evolution histories. To date\footnote{According to the NASA Exoplanet Archive \citep{NASA_EXO_Archive_Akeson_2013} as of December 28, 2024}, 63 sub-Saturn exoplanets have been detected with masses and radii constraints better than 20\%. Among these, 37 are classified as hot sub-Saturns, with orbital periods shorter than 10 days. Notably, only 4 of these hot sub-Saturns are found in multiple star systems.

In this work, we present the discovery and characterization of the sub-Saturn exoplanet TOI-6038~A~b, located at the outer edge of the newly identified Neptunian ridge. The presence of a wide binary companion in the system makes this discovery particularly valuable for understanding the role of secular perturbations in HEM. In Sec. \ref{sec:obs}, we describe the TESS photometry, PARAS-2 high-resolution spectroscopy, PRL 2.5 speckle imaging observations, and their data analysis procedures. The system characterization and global modeling are discussed in Sec. \ref{sec:analysis}. The dynamical history of TOI-6038~A~b and its contextualization in the current exoplanet population is discussed in Sec. \ref{sec:discussion}. We summarize our findings in Sec. \ref{sec:summary}.


\section{Observations} \label{sec:obs}

\subsection{TESS photometry} \label{sec:tess}

\begin{figure*}[t!]
\centering
\includegraphics[width=0.42\textwidth]{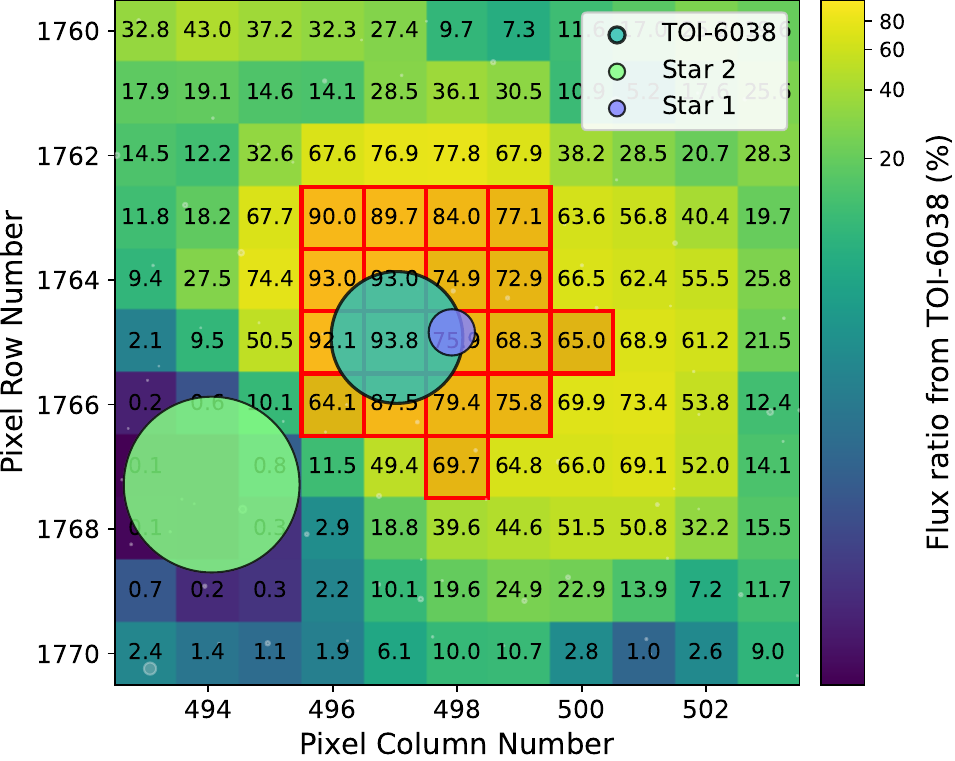}
\hspace{0.7cm}
\includegraphics[width=0.42\textwidth]{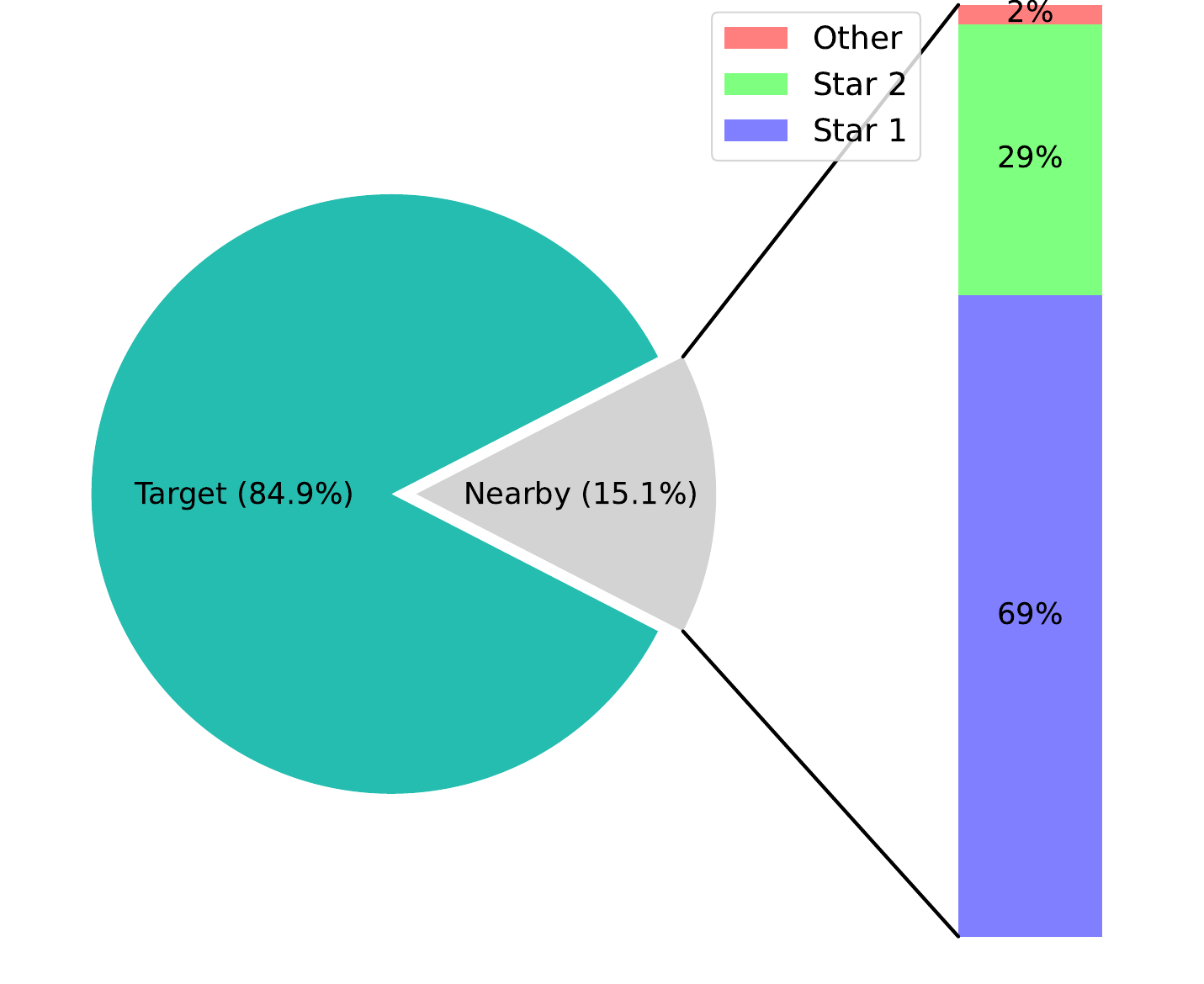}
    \caption{Nearby sources contaminating the TOI-6038~A aperture. Left: TPF-shaped heatmap with the pixel-by-pixel flux fraction from TOI-6038~A in S58. The red grid is the SPOC aperture. The pixel scale is 21 arcsec $\rm pixel^{-1}$. The two sources that most contribute to the aperture flux (Star 1: TIC 194736419 and Star 2: TIC 194736424) are highlighted in purple and green. Disk areas scale with the emitted fluxes. Right: Flux contributions to the SPOC aperture from the target and most contaminant stars. These plots were generated through \texttt{TESS-cont} \citep{CastroGonzalez2024b}.}
    \label{fig:TESS-cont}
\end{figure*}

\begin{figure*}[t!]
\centering
\includegraphics[width=0.97\textwidth]{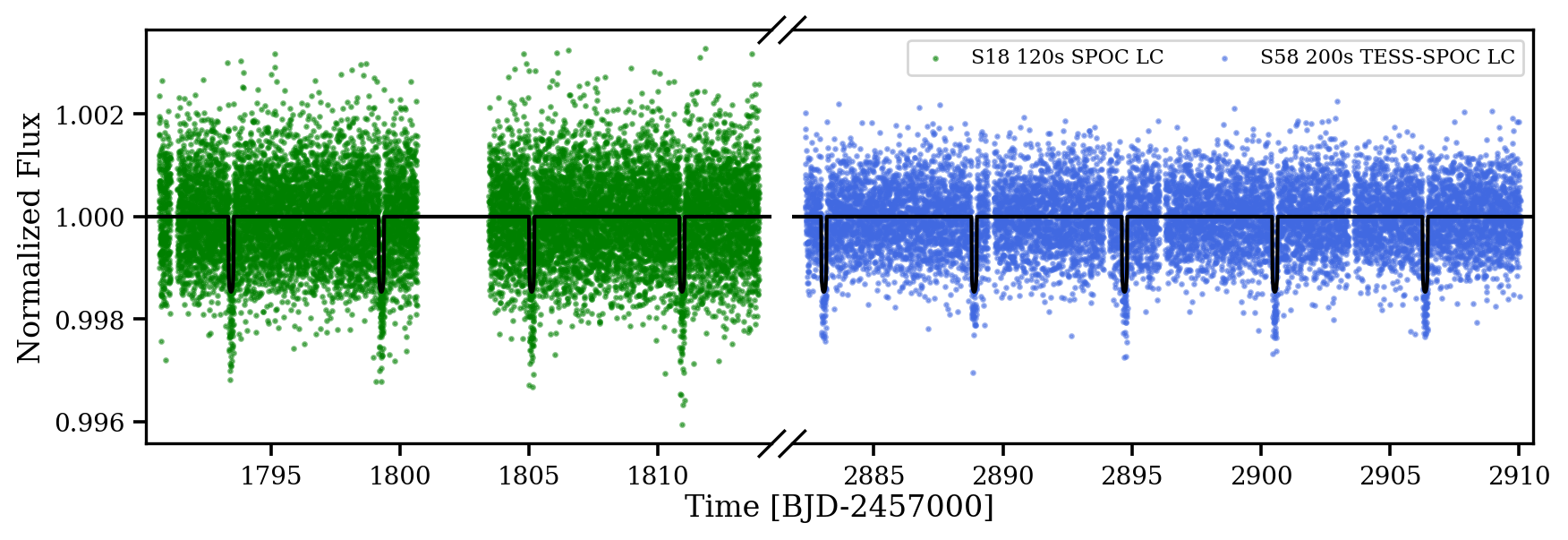}
\includegraphics[width=0.96\textwidth]{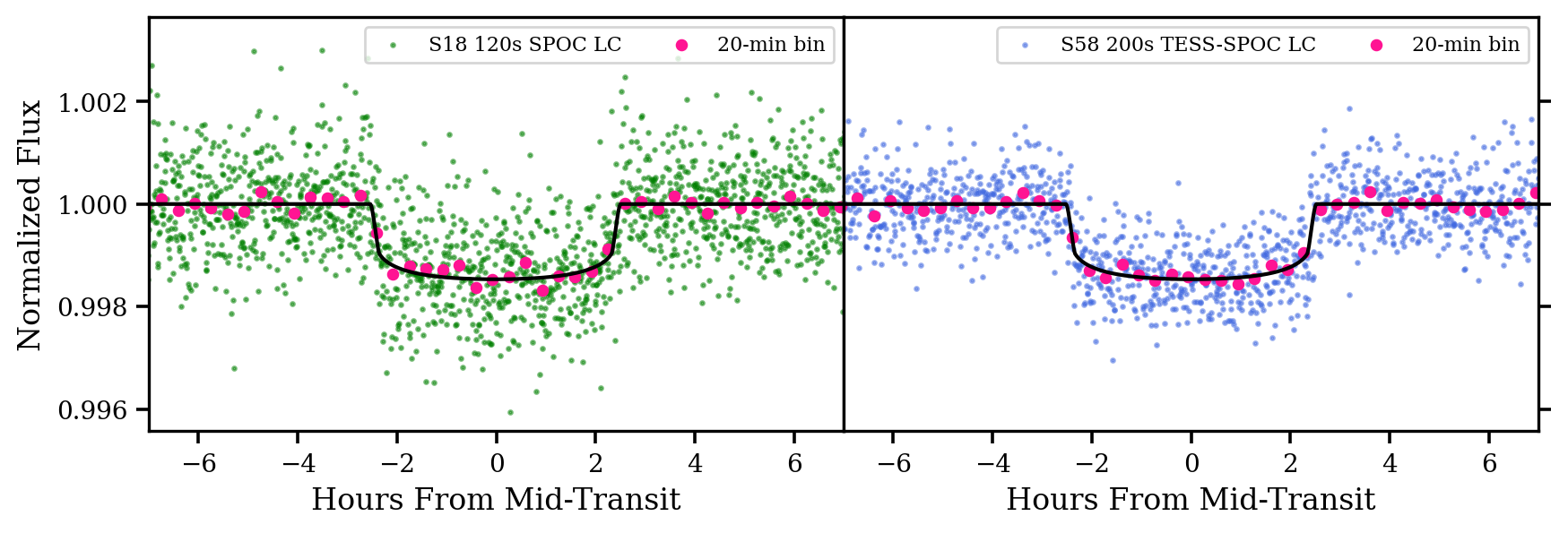}
    \caption{Detrended and normalized TESS light curves (LCs) of TOI-6038~A from S18 and S58 are shown with green and blue points, respectively. The full LC is displayed in the upper panel, while the phase-folded LC from both sectors is shown in the lower panel, with pink dots representing 20-minute binned data points. The black solid line in both panels represents the best-fit transit model for TOI-6038~A~b (see Sec. \ref{sec:global_modeling}).}
    \label{fig:tesslc}
\end{figure*}

TOI-6038~A (TIC 194736418) was observed by TESS in Sectors 18 and 58. The Sector 18 (S18) observations were conducted between November 3 and November 27, 2019, while the Sector 58 (S58) observations took place from October 29 to November 26, 2022. In both sectors, TOI-6038~A was observed using CCD~1 of Camera~1. The target is also scheduled for observation in Sector 86 in December 2024. The light curve (LC) from the S18 target pixel file (TPF) was produced by the Science Processing Operations Center \citep[SPOC;][]{spoc} pipeline in 120-second cadence mode. The LCs were also generated from TESS Full Frame Images (FFIs) using the TESS-SPOC pipeline \citep{tess-spoc} and the Quick Look Pipeline \citep[QLP;][]{qlp1,qlp2}, with a cadence of 1800 seconds for S18 and 200 seconds for S58. All LCs for TOI-6038~A are publicly available through the Mikulski Archive for Space Telescopes (MAST)\footnote{\url{https://mast.stsci.edu/portal/Mashup/Clients/Mast/Portal.html}}.

We searched for nearby contaminant stars through the TESS contamination tool \texttt{TESS-cont}\footnote{Available at \url{https://github.com/castro-gzlz/TESS-cont}.} \citep{CastroGonzalez2024b}. In brief, the package searches for all nearby \textit{Gaia} DR2/DR3 sources \citep{gaiaedr3} and builds their Pixel Response Functions (PRFs) to estimate the flux distribution across the TPF of the target star. The PRFs are then scaled to the stellar relative fluxes and placed in a TPF-shaped array to compute the flux contribution from each source. In Fig.~\ref{fig:TESS-cont}, we show the S58 \texttt{TESS-cont} output for TOI-6038~A. As we can see, TOI-6038~A is located in a fairly uncrowded field, but there are two relatively nearby sources that contribute $15\%$ of the aperture flux. Star~1 (TIC 194736419; $G$ = 12.01 mag) is located inside the aperture, at a projected separation of 17.87\arcsec from TOI-6038~A, and with a total flux contribution of 10.4$\%$. Star~2 (TIC 194736424; $G$ = 9.10 mag) is located outside the aperture, at a projected separation of 78.6\arcsec from TOI-6038~A, and with a total flux contribution of 4.4$\%$. Interestingly, Star~1 is identified as a binary companion to TOI-6038~A (see Sect. \ref{sec:binary_system} for further details). The flux contributions presented above together with the locations of the sources imply that, based on TESS data alone, Star~1 cannot be discarded as being the origin of the observed transit signal \citep[e.g.][]{2018AJ....156..277L,2021MNRAS.508..195D,2022MNRAS.509.1075C}. This is very common for TESS candidates given the large apertures and pixel sizes. To confirm the origin of the transit signal, it is necessary to complement the TESS data with follow-up ground-based observations of higher spatial resolutions (see Sects.~\ref{sec:speckle} and \ref{sec:paras2_obs}).

For the photometric analysis, we used the S18 and S58 Pre-search Data Conditioning Simple Aperture Photometry \citep[PDCSAP;][]{smith_2012} flux time series, generated by the SPOC and TESS-SPOC pipelines, respectively. The SPOC pipelines calculate the optimal apertures for extracting light curves from each target and estimate contamination from nearby stars within the same aperture to correct for crowding. We used \texttt{citlalicue} \citep{citlalicue} to detrend and normalize the light curves, employing Gaussian process regression to remove uncorrected systematics and low-frequency stellar variability. Since the PDCSAP light curves are already corrected for dilution, we did not include a dilution factor for TESS in the fit. The final de-trended and normalized light curves are shown in Fig.~\ref{fig:tesslc}. In total, nine transits were observed, four in S18 and five in S58.

\subsection{Speckle imaging with PRL 2.5m telescope}\label{sec:speckle}

\begin{figure}[t!]
    \centering
    \includegraphics[width=0.5\textwidth]{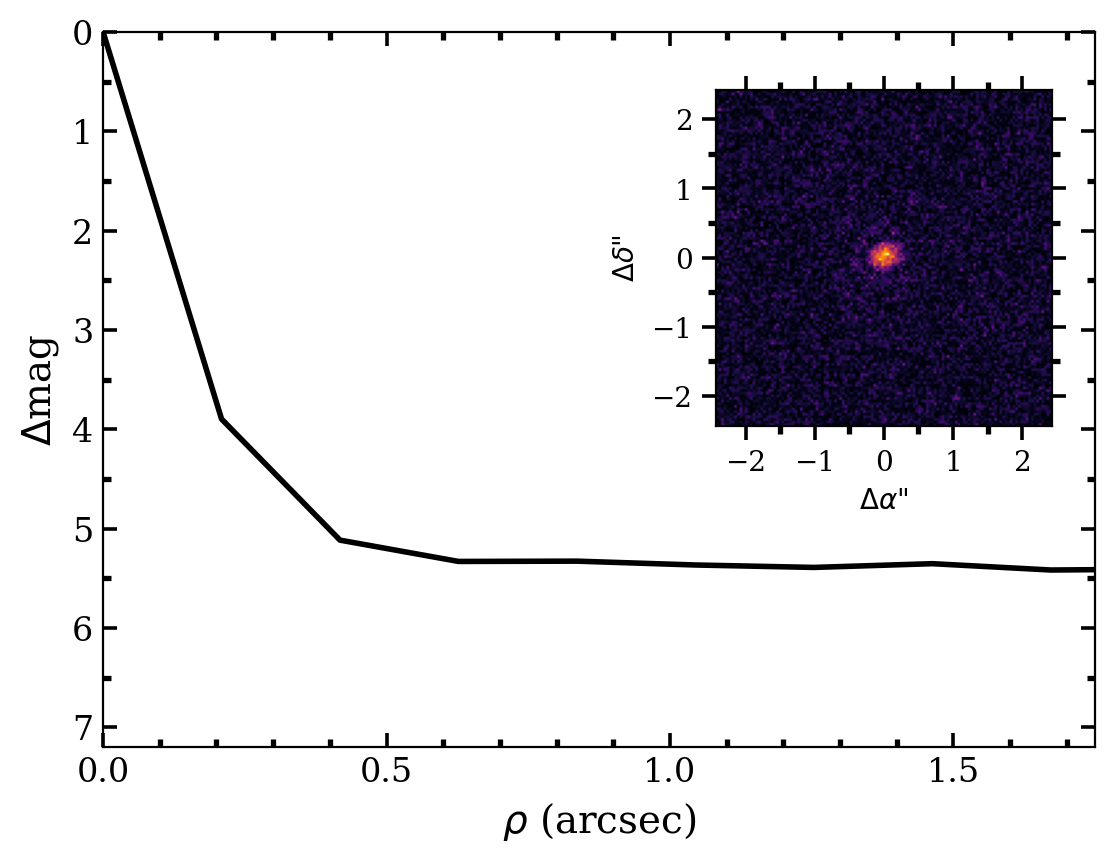}
    \caption{5$\sigma$ contrast curve in the V band for TOI-6038~A, obtained using the speckle imager on the PRL 2.5m telescope. The speckle ACF is shown as an inset. No stellar companions are detected.}
    \label{fig:speckle}
\end{figure}

Speckle imaging of TOI-6038~A was performed in January 2024 using the speckle imager on the 2.5m telescope at the PRL Observatory in Gurusikhar, Mount Abu, Rajasthan, India. The speckle imager is equipped with a TRIUS PRO-814 CCD detector (model ICX814AL)\footnote{\url{https://www.sxccd.com/product/trius-sx814//}} featuring a resolution of 3388 × 2712 pixels, with individual pixel dimensions of 3.69 $\mu m$ × 3.69 $\mu m$. The field of view (FOV) is 2.15$\arcmin$ × 1.7$\arcmin$, and the pixel scale is 38 mas/pixel. Observations were conducted using the Bessel-V filter. Around 5000 speckle frames, each with a 10 ms exposure time, were captured under favorable sky conditions, with an average seeing around 1$\arcsec$. Data reduction was carried out using a custom pipeline written in {\tt PYTHON}, based on procedures similar to those described by \cite{2020ziegler}  and \cite{2018tokovinin}. The analysis involved processing the power spectral density function, constructing the autocorrelation function (ACF), and estimating the 5$\sigma$ contrast. An overview of the steps involved is explained in \cite{toi6651}.

The derived contrast limits for TOI-6038~A are $\Delta V = 4.10$ at 0.25\arcsec and $\Delta V = 5.36$ at 1\arcsec. No stellar companions are detected. Figure \ref{fig:speckle} presents the 5$\sigma$ contrast curve alongside the ACF for TOI-6038~A. In addition, the re-normalized unit weight error (RUWE) from \textit{Gaia} DR3 \citep{gaia2023} is a robust metric for testing unresolved multiplicity in \textit{Gaia} sources. The RUWE for TOI-6038~A is $\approx 0.93$, suggesting that the single-star model provides a good fit to the astrometric observations.

\subsection{PARAS-2 radial velocity observations}{\label{sec:paras2_obs}}

PARAS-2 is a newly installed high-resolution spectrograph ($R = 107,000$) attached to the 2.5m telescope at the PRL Observatory \citep{paras2_design, prl2.5m_and_paras2}. It uses a simultaneous Uranium-Argon hollow cathode lamp for wavelength calibration and operates in the wavelength range of 3800–6900 \AA. The instrument has demonstrated an RV stability of 2.65 m s$^{-1}$ over a 35-day period on a standard star \citep{toi6651}. The RVs are calculated by cross-correlating the wavelength-calibrated spectra with a template spectrum \citep{1996A&AS..119..373B}. Details of the data reduction and pipeline analysis are provided in \cite{toi6651}.
We observed TOI-6038~A using PARAS-2 on three separate occasions. The first set of observations took place from January 31 to February 3, 2023, during which we collected four good-quality spectra. The second set occurred between November 16 and November 24, 2023, yielding six spectra. In the final set, we collected nineteen spectra from December 28, 2023 to January 24, 2024. In total, we acquired twenty-nine spectra, with each spectrum observed for 3600 seconds.
We would also like to mention that fourteen additional spectra were excluded from the RV analysis due to poor seeing conditions at the time of observation. The details of the RV measurements from the twenty-nine good spectra are presented in Table \ref{tab:rv_table}. The errors in these measurements range from 4.6 to 21.7 m s$^{-1}$, with a median error of 8.3 m s$^{-1}$. Notably, the errors from the first set of data (January 31 to February 3, 2023) are higher than those from subsequent observations. This increased error can be attributed to the condition of the telescope's mirror, which was recoated in October 2023. After the recoating, we obtained data with significantly improved S/N. The median S/N per pixel for the spectra we observed is 26.

PARAS-2 is a fiber-fed spectrograph with a 75-micron core fiber, corresponding to a FOV of 1.5\arcsec on the sky \citep{paras2_design}. As discussed in Sec. \ref{sec:speckle}, TOI-6038~A has no unresolved binary companions, and the nearest star, TOI-6038~B, is 18\arcsec away, which is too distant to contaminate the fiber. Thus, the observed RV variations can be attributed solely to TOI-6038~A without any contamination from nearby stars.


\section{Result and analysis} \label{sec:analysis}

\renewcommand{\arraystretch}{1.1}
\begin{table*}[t!]
\small
\centering
\caption{Basic parameters of TOI-6038 binary star system.}
\label{tab:star_table}
\begin{tabular}{llll}
\hline
\hline
\noalign{\smallskip}
Parameter&TOI-6038~A&TOI-6038~B&Ref.\\
\noalign{\smallskip}
\hline
\noalign{\smallskip}
Identifiers:\\
TIC&194736418&194736419&(1)\\
\textit{Gaia}DR3& 237640407346340736 & 237640407346341376 &(2)\\
2MASS& J03263773+4049029& J03263617+4049006 &(3)\\
SUPERWIDE & SWB70395 & -- & (4)\\
\hline
\noalign{\smallskip}
Astrometry:\\
$\alpha_{J2000}$ & 03:26:37.731 & 03:26:36.168 & (2)\\
$\delta_{J2000}$ & +40:49:02.79 & +40:49:00.62 & (2)\\
$\mu_{\alpha}$ (mas yr$^{-1}$) &  10.184 $\pm$ 0.020 & 10.712 $\pm$ 0.016 & (2)\\
$\mu_{\delta}$ (mas yr$^{-1}$) & -40.002 $\pm$ 0.022 & 	-40.023 $\pm$ 0.015 & (2)\\
$\varpi${\textdagger} (mas) & 5.644 $\pm$ 0.020 & 5.630 $\pm$ 0.018 & (2)\\
$d$ (pc) & $177.20\pm0.63$ & $177.62 \pm 0.58 $ & (2)\\
Ang. Sep. ($\arcsec$) & -- & 17.87 & (4)\\
Proj. Sep. (AU) & -- & 3217 & (4)\\
\hline
\noalign{\smallskip}
Photometry{\textdagger}:\\
$T$   & 9.3229 $\pm$ 0.0061 & 11.4394 $\pm$ 0.0061 & (1)\\
$G$   & 9.734  $\pm$ 0.020 & 12.003 $\pm$ 0.020 & (2)\\
$G_{BP}$   & 10.034 $\pm$ 0.020 & 12.466 $\pm$ 0.020 & (2)\\
$G_{RP}$   & 9.268 $\pm$ 0.020 & 11.380 $\pm$ 0.020 & (2)\\
$J$   & 8.727 $\pm$ 0.020 & 10.607 $\pm$ 0.021 & (3)\\
$H$   & 8.502 $\pm$ 0.021 & 10.205 $\pm$ 0.021 & (3)\\
$K_{S}$ & 8.445 $\pm$ 0.020 & 10.111 $\pm$ 0.020 & (3)\\
$W1$  &  8.406 $\pm$ 0.030 & 10.052 $\pm$ 0.030 & (5)\\
$W2$  &  8.453 $\pm$ 0.030 & 10.111 $\pm$ 0.030 & (5)\\
$W3$  &  8.413 $\pm$ 0.030 & 10.118 $\pm$ 0.062 & (5)\\
\hline
\noalign{\smallskip}
\multicolumn{4}{l}{Spectroscopic parameters from TRES SPC:}\\
$T_{\rm eff}$ (K) &$6044\pm65$ & -- & (6)\\
$\log{g}$ (cgs) &$4.11\pm0.11$ & -- & (6)\\
$[{\rm Fe/H}]$ (dex) &$0.097\pm0.085$ & -- & (6)\\
$v\sin{i_*}$ ($km \ s^{-1}$) & $7.94 \pm 0.50$ & -- & (6)\\
\hline
\noalign{\smallskip}
\multicolumn{4}{l}{Derived stellar parameters{\textdaggerdbl}:}\\
$M_*$ (\msun) &$1.291^{+0.066}_{-0.060}$ & $0.86\pm0.10$ & (6,7)\\
$R_*$ (\rsun) &$1.648^{+0.045}_{-0.035}$ & $0.896\pm0.054$ & (6,7)\\
$L_*$ (\lsun) &$3.41^{+0.28}_{-0.26}$ & $0.493\pm0.024$ & (6,7)\\
$\rho_*$ (cgs) &$0.412^{+0.017}_{-0.033}$ & $1.68\pm0.39$ & (6,7)\\
$T_{\rm eff}$ (K) &$6110\pm100$ & $5109\pm117$ & (6,7)\\
$\log{g}$ (cgs) &$4.118^{+0.015}_{-0.025}$ & $4.467\pm0.080$ & (6,7)\\
$[{\rm Fe/H}]$ (dex) &$0.124^{+0.079}_{-0.077}$ & -- & (6)\\
$Age$ (Gyr) &$3.65^{+0.92}_{-0.85}$ & -- & (6)\\
$A_V$ (mag) &$0.209^{+0.096}_{-0.098}$ & -- & (6)\\
\noalign{\smallskip}
\hline
\noalign{\smallskip}
\end{tabular}\\
\footnotesize{\textbf{Notes.} {\textdagger}A systematic error floor has been applied to the uncertainties.\\{\textdaggerdbl}Parameters for TOI-6038~A are obtained from the global modeling (Sect. \ref{sec:global_modeling}), while those for TOI-6038~B are sourced from TICv8.2.\\
\textbf{References:} (1) \cite{2018AJ....156..102S},  (2) \cite{gaia2023}, (3) \cite{JHK}, (4) \cite{superwide_2020}, 
(5) \cite{ALLWISE}, (6) This work, (7) \cite{ticv8.2}}
\end{table*}

\subsection{Stellar parameters from spectral synthesis}\label{sec:spec_synthesis}

We acquired four spectra of TOI-6038~A with SNRs ranging from 41 to 101, covering wavelengths from 3850 to 9096 \AA, using the TRES spectrograph ($R=44,000$) on the FLWO 1.5m telescope in February and November 2023. The stellar parameters were derived using the Stellar Parameter Classification tool \citep[SPC;][]{buchhave2010, buchhave2012, buchhave2014}, which cross-correlates with a grid of synthetic spectra based on Kurucz atmospheric models \citep{kurucz1992}. The weighted average of the stellar parameters resulted in an effective temperature ($T_{\rm eff}$) of $6044 \pm 65$ K, a surface gravity ($\log{g}$) of $4.11 \pm 0.11$, a metallicity ($[{\rm Fe/H}]$) of $0.097 \pm 0.085$ dex, and a rotational velocity ($v\sin{i}_*$) of $7.94 \pm 0.50$ km s$^{-1}$, as listed in Table \ref{tab:star_table}. We used these parameters as prior constraints and to provide initial values in the global modeling using EXOFASTv2 \citep{exofast} for a comprehensive stellar characterization (see Sec. \ref{sec:global_modeling}).

\subsection{TOI-6038 binary star system} \label{sec:binary_system}

Binary star systems are generally classified into two categories: tight binary system, where the distance between the components is less than 1000 AU, and wide binary system (WBS), with a mutual distance greater than 1000 AU. TOI-6038~A has been identified as a primary star in a WBS within the SUPERWIDE catalog \citep{superwide_2020}, with a 99.77\% probability of being a real binary. We refer to its binary companion as TOI-6038~B (i.e. TIC 194736419, and Star 1 in Fig. \ref{fig:TESS-cont}). The angular separation between the two stars is approximately $17.87\arcsec$, and the difference in $G$ band magnitude is $\Delta G \approx 2.281$. The projected physical separation between the stars is about 3217 AU. RV difference between the two stars is 1.83 km s$^{-1}$, calculated from \textit{Gaia} RVs \citep{gaia2023}. The astrometry, photometry, and basic stellar properties for both components are given in Table \ref{tab:star_table}.

TOI-6038~B is an early K-type dwarf with $T_\mathrm{eff} = 5109 \pm 117$ K, $\log{g} = 4.467 \pm 0.080$, $M_* = 0.86 \pm 0.10 \ \msun$, and $R_* = 0.896 \pm 0.054 \ \rsun$, as listed in the TESS Input Catalog \citep[TIC v8.2;][]{ticv8.2}. These stellar parameters are also consistent with those derived using Gaia's BP/RP spectra through the General Stellar Parametrizer from Photometry \citep[GSP-Phot;][]{GSP-phot_23}. The lower limit on the orbital period of this binary system, assuming a circular orbit, is approximately 124,478 years. TOI-6038 is only the fifth known multiple star system to host a hot sub-Saturn, among those with planet masses and radii constraints better than 20\%.

\subsection{Galactic kinematics}

The Galactic space velocity components $UVW$ for TOI-6038~A in the barycentric frame were estimated using the \texttt{gal$\_$uvw}\footnote{\url{https://pyastronomy.readthedocs.io/en/latest/pyaslDoc/aslDoc/gal_uvw.html}} function. These velocity components are defined as positive in the directions of the Galactic center, Galactic rotation, and the north Galactic pole, with values of 3.360, -31.638, and -19.018 $km \ s^{-1}$, respectively. These velocities are also expressed relative to the local standard of rest (LSR), employing the solar velocities from \cite{Sch2010}, which are 14.460, -19.398, and -11.768 $km \ s^{-1}$, respectively. Our analysis indicates that TOI-6038~A is associated with the thin disk population \citep{Leggett1992, Bensby2014}. Furthermore, the BANYAN $\Sigma$ algorithm, which determines cluster membership probabilities based on sky coordinates, parallaxes, proper motions, and radial velocities (RVs), classifies TOI-6038~A as a field star with over 99$\%$ probability, showing no association with known young stellar clusters \citep{Gagne2018}.

\subsection{Rotational period of the star} \label{sec:rot_per}

\begin{figure*}[t!]
\centering
\includegraphics[width=\textwidth]{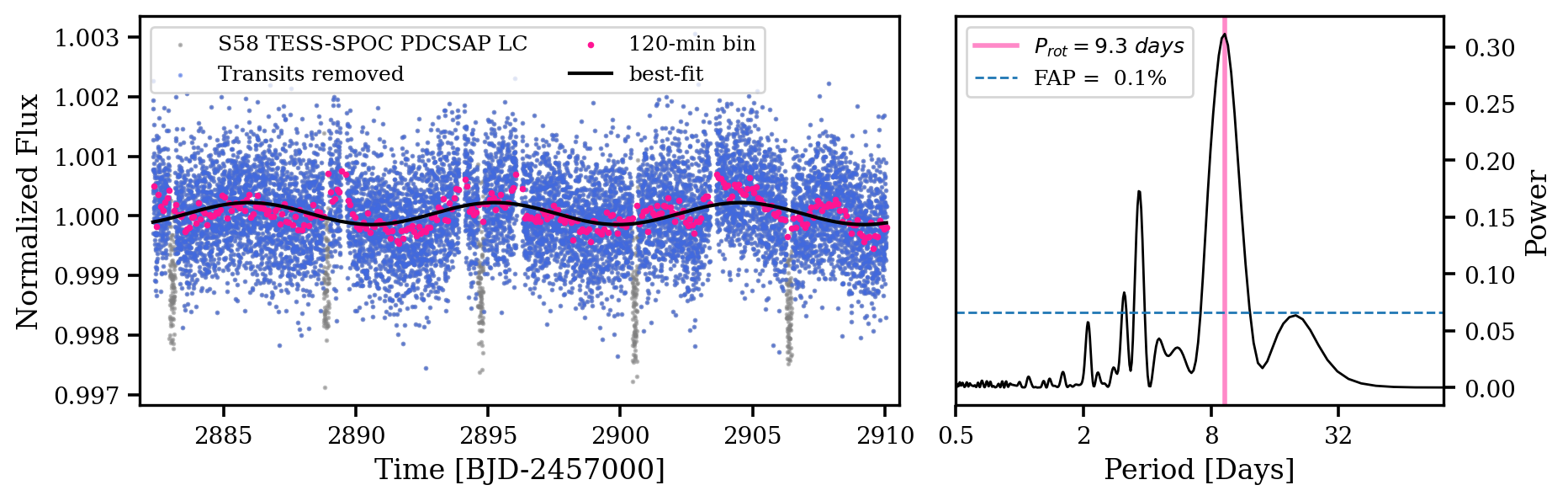}
    \caption{Left: Normalized TESS-SPOC PDCSAP light curve (LC) from S58, shown with grey points (full LC) and blue points (with transits removed). The pink dots represent 120-minute binned data points, while the solid black line shows the best-fit sinusoidal signal. Right: GLS periodogram of the LC. The vertical pink line marks the most significant period ($9.29 \pm 0.25$ days), and the horizontal dashed blue lines indicate the 0.1\% FAP level.}
    \label{fig:periodogram_TESS}
\end{figure*}

The upper limit on the stellar rotation period, assuming the star is observed equator-on, is calculated to be $11.2$ days based on the projected rotational velocity and stellar radius given in Table \ref{tab:star_table}. We also analyzed the photometric variability in the TESS-SPOC PDCSAP light curves (LCs). If the star is sufficiently active, the signal from stellar rotation variations is typically preserved in the PDCSAP data for stars with rotation periods shorter than 13 days. In the S58 data, we identified significant sinusoidal modulation, as shown in Fig. \ref{fig:periodogram_TESS}. To determine the periodicity, we first removed the in-transit data from the LC and binned it to 120-minute intervals. We then employed the Generalized Lomb-Scargle (GLS) periodogram \citep{periodogram}. The most prominent peak in the periodogram corresponds to a period of $9.29 \pm 0.25$ days (see Fig. \ref{fig:periodogram_TESS}). It is consistent with the stellar rotation period derived from the rotational velocity and radius. The small difference in values can be explained by the possibility that the star is not perfectly equator-on, with the discrepancy corresponding to a stellar inclination angle of 62$^{\circ}$. However, it is important to note that the LC is contaminated by nearby sources (accounting for $\approx15\%$ of the aperture flux), which raises the possibility that the detected signal may originate from a nearby star. Nonetheless, we remain open to the chance that the observed modulation could reflect the stellar rotation signal. We also inspected the LC from S18, but no such significant periodic signal was detected, possibly lost during data processing used to generate the PDCSAP-reduced data.

\subsection{Periodogram analysis of the RVs} \label{sec:periodogram}

\begin{figure}[t!]
    \centering
    \includegraphics[width=0.5\textwidth]{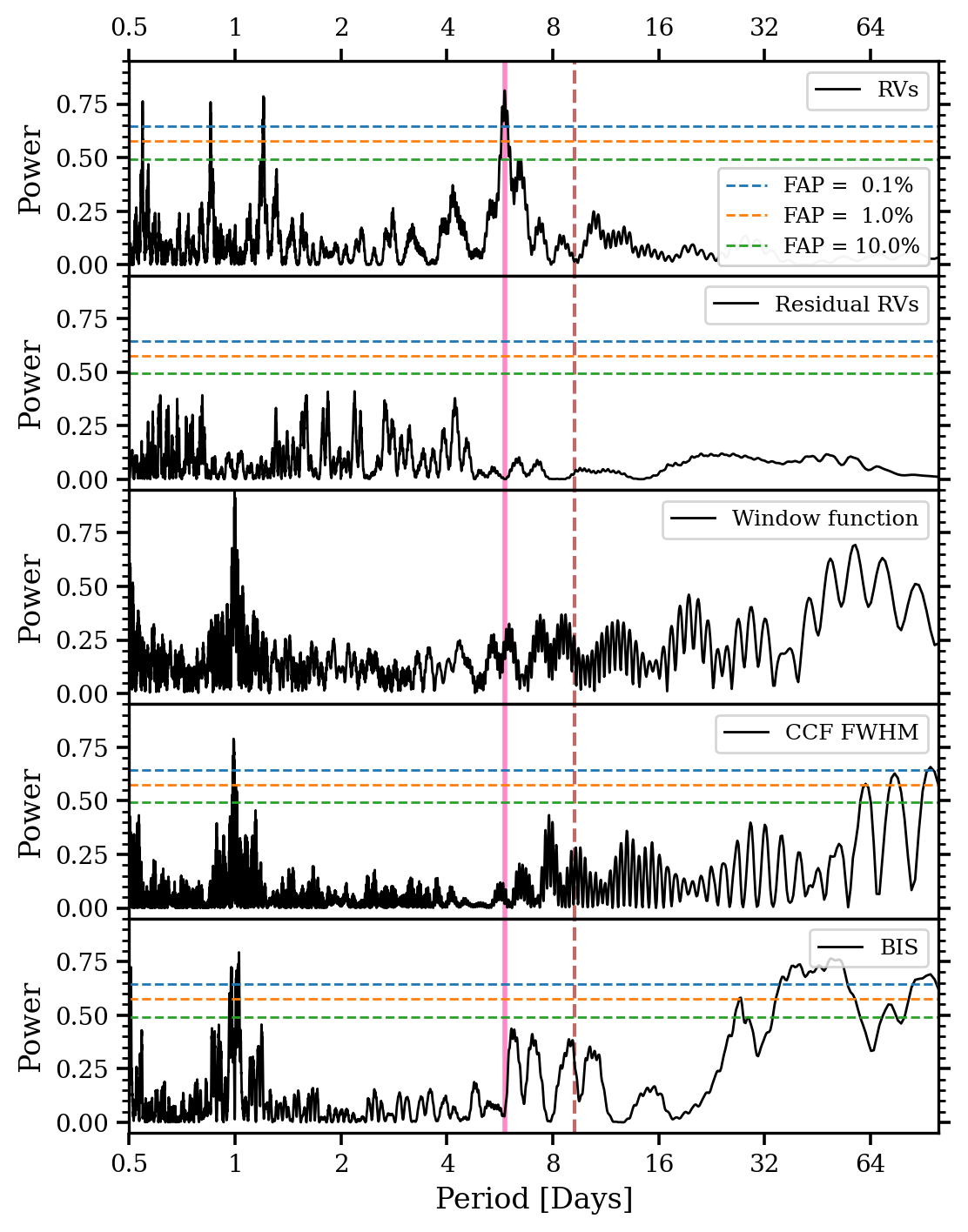}
    \caption{GLS periodogram of RVs, residual RVs, window function, CCF FWHM, and bisector span, shown in panels 1–5 (top to bottom), respectively. The vertical pink line marks the most significant period ($5.832 \pm 0.010$ days) found in the RVs, which is consistent with the period found by TESS. The rotational period of the star ($9.29$ days) is marked by the vertical brown dashed line. The FAP levels of 0.1\%, 1.0\%, and 10.0\% are indicated by horizontal dashed lines in blue, orange, and green, respectively.}
    \label{fig:periodogram}
\end{figure}

We used GLS periodogram \citep{periodogram} to identify sinusoidal periodic signals in the RV data observed with PARAS-2, independently of the orbital period found in TESS, as shown in panel 1 of Fig. \ref{fig:periodogram}. The most significant peak in the GLS periodogram corresponds to a period of $5.832 \pm 0.010$ days, consistent with the period detected by TESS ($P \approx 5.827$ days). The window function, which represents the Fourier transform of the time-sampling pattern of the observational data, was also analyzed (panel 3, Fig. \ref{fig:periodogram}). Periodicities in the window function can interact with the true signal, introducing aliases in the periodogram. The other peaks in RV periodogram above the 0.1\% false alarm probability (FAP) level are observed at approximate periods of 0.55, 0.85, and 1.20 days. We then subtracted the best-fit sine wave (5.832 days) from the RV data and recalculated the periodogram (panel 2, Fig. \ref{fig:periodogram}). The residual periodogram revealed no significant peaks, suggesting that the additional peaks are likely aliases of the primary signal. The RVs of TOI-6038~A are not affected by any nearby contamination (see Sec. \ref{sec:paras2_obs}), while the light curve has flux contributions from nearby sources (see Sec. \ref{sec:tess}). The detection of the same orbital period in both the RVs and TESS photometric data, with the two in-phase (see Sec. \ref{sec:global_modeling}), suggests that the transit signals in the light curve originate from TOI-6038~A rather than from any other nearby star.

RV variations can also be induced by stellar activity, such as star spots and/or faculae \citep{2012A&A...545A.109B,2014ApJ...796..132D,Fischer_2016,2018A&A...614A..76J,2019MNRAS.489.2555D,2023A&A...670A..24S}, which have been detected in combination with planetary signals \citep[e.g.][]{2021A&A...654A..60L,2023A&A...675A..52C,2023A&A...679A..33D,2024A&A...690A..79G} and have the potential of mimicking them \citep[e.g.][]{2008A&A...489L...9H,2009A&A...493..645F,2014A&A...566A..35S}. In addition to RVs from PARAS-2 spectra, we have measured the full width at half maximum (FWHM) and the bisector inverse slope (BIS) of the cross-correlation function (CCF), both of which serve as stellar activity indicators \citep{Bisector1}. The GLS periodograms for the CCF FWHM and BIS (panel 4 \& 5, Fig. \ref{fig:periodogram}) show no significant peaks around the best-fit RV period of 5.83 days, and not additional relevant peaks pop up. We can thus now conclude that the periodic signal detected in the PARAS-2 RV data is not caused by stellar activity but rather is due to a planet orbiting around TOI-6038~A previously identified as a TESS candidate.

\subsection{Planetary system parameters from global modeling} \label{sec:global_modeling}

\begin{figure}[t!]
    \centering
    \includegraphics[width=0.5\textwidth]{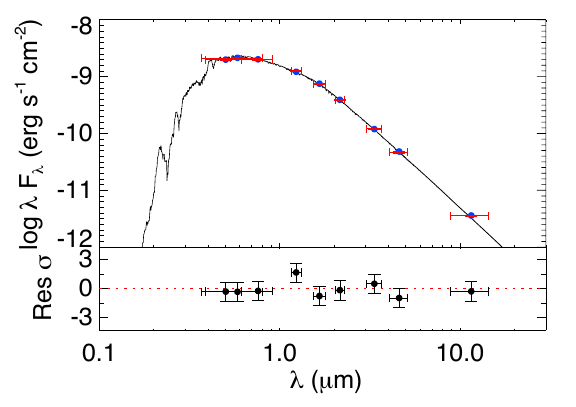}
    \caption{SED of TOI-6038~A, with red symbols representing the observed photometric measurements and horizontal bars indicating the effective width of the passbands. The blue points represent the model fluxes, and the residuals are displayed in the lower panel.}
    \label{fig:sed}
\end{figure}

\begin{figure}[t!]
    \centering
    \includegraphics[width=0.5\textwidth]{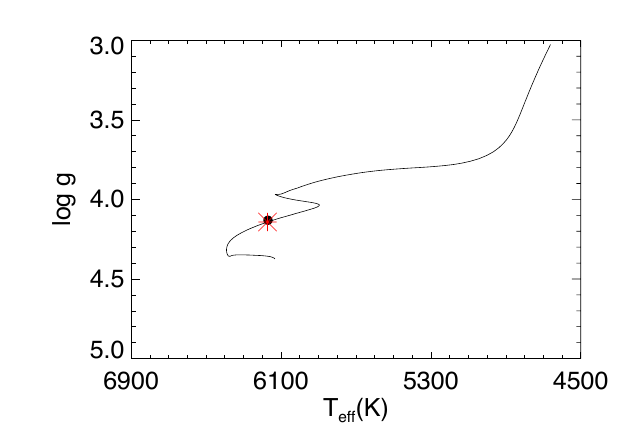}
    \caption{MIST evolutionary track for TOI-6038~A shown as a solid black line. The black point indicates the $T_\mathrm{eff}$ and $\log{g}$, while the red asterisk denotes the current age of TOI-6038~A.}
    \label{fig:mist}
\end{figure}

We used EXOFASTv2 \citep[][hereafter EF2]{exofast} to determine the stellar and planetary parameters of the TOI-6038~A system. EF2 is an exoplanet-fitting suite written in {\tt IDL} that can simultaneously model both RV and transit data. Additionally, it provides stellar parameters by fitting the spectral energy distribution \citep[SED;][]{SED1} and using MIST stellar evolutionary models \citep{mist_choi, mist_dotter}. The software operates based on the Markov Chain Monte Carlo (MCMC) algorithm, and convergence is achieved when the Gelman-Rubin statistic is less than 1.01, and the number of independent draws exceeds 1000 \citep{Gelman_rubin2006}.

The combination of SED, isochrones, and transit data allows for a precise determination of the stellar mass, radius, and surface gravity \citep{Torres2008}. We applied Gaussian priors on $T_{\text{eff}}$ and $[\text{Fe}/\text{H}]$ based on the spectroscopic parameters, and on the parallax derived from \textit{Gaia} DR3 \citep{gaia2023}. The parallax (5.612 mas) was corrected by adding 0.031 mas, and its uncertainty (0.0172 mas) was combined in quadrature with 0.01 mas to account for residual systematics, following \cite{Lindegren_2021}. For SED fitting, we used photometric magnitudes from the \textit{Gaia} $G$, $G_{BP}$, and $G_{RP}$ bands, along with the 2MASS $J$, $H$, and $K_S$ bands, and the WISE $W1$, $W2$, and $W3$ bands, as listed in Table \ref{tab:star_table}. We excluded the $W4$ band due to potential contamination from the companion star, TOI-6038~B. To account for systematic errors in the absolute photometry, the uncertainties in these magnitudes were inflated following the method outlined in \citet{Lindegren_2021}. Additionally, we applied a uniform prior to enforce an upper limit on the V-band extinction, using the \cite{extinction} dust maps at the location of TOI-6038~A.

We performed a global model fit with EF2, considering both eccentric and circular orbits. Initial values for the orbital period ($P$) and central transit time ($T_C$) are taken from the SPOC data validation report. The RV model is generated using Kepler’s laws, while the transit model is created using the approach of \cite{Mandel2002} and \cite{Agol_2020}, with limb-darkening parameters constrained by \cite{Claret} and \cite{Claret_tess}. The value of eccentricity from the eccentric orbit fit is $e=0.045^{+0.056}_{-0.032}$. To compare the eccentric and circular fits, we computed the Bayesian Information Criterion \citep[BIC;][]{BIC} and Akaike Information Criterion \citep[AIC;][]{AIC}. The difference in AIC between the models is 5.62, while the difference in BIC is 21.94, strongly favoring the circular orbit model. Consequently, we adopt the circular model for further analysis.

Both the eccentric and circular EF2 fits revealed slight bimodality in the posterior distributions of stellar mass and age. This bimodality is not unprecedented; it is a known feature in certain regions of the MIST models \citep[e.g.][]{Carmichael_2021,toi6651}. The two peaks in mass are found at 1.145 and 1.291 \msun. We split the distribution at the local minimum of 1.170 {\msun} and extracted both solutions, ultimately adopting the high-mass solution due to its higher relative probability (87\%). All reported parameters and subsequent analyses are based on this solution.

\begin{figure*}[t!]
    \centering
    \includegraphics[width=0.48\textwidth]{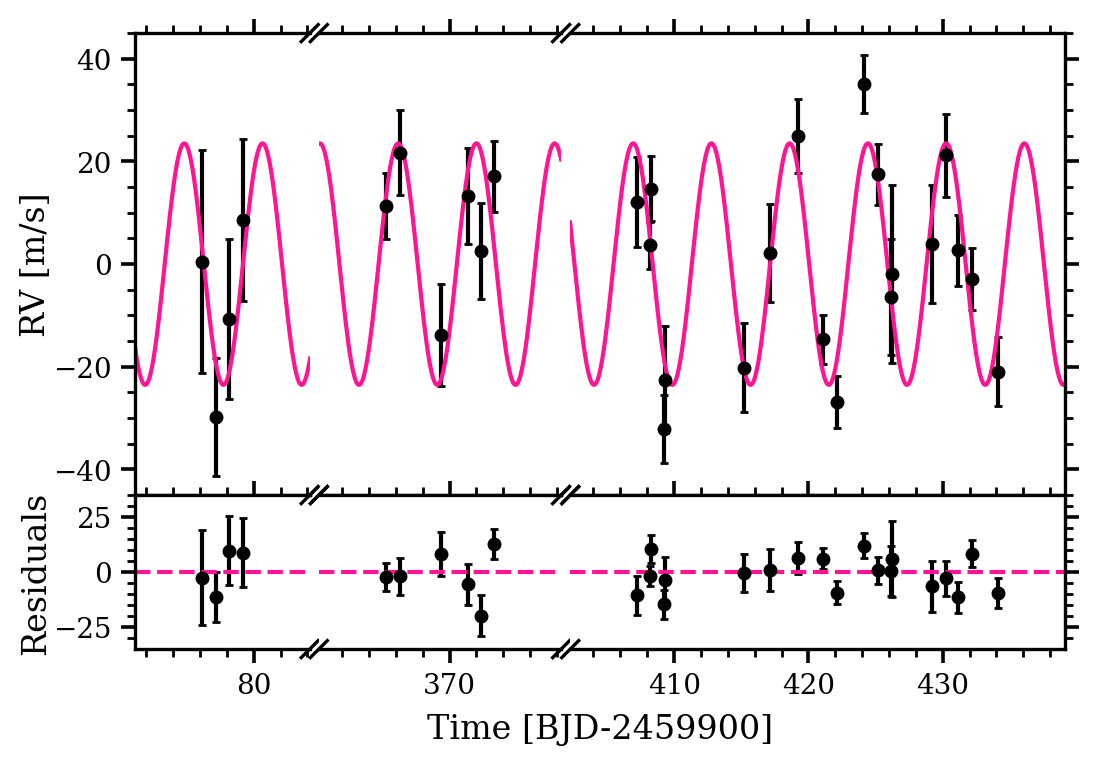}
    \hspace{0.3cm}
    \includegraphics[width=0.48\textwidth]{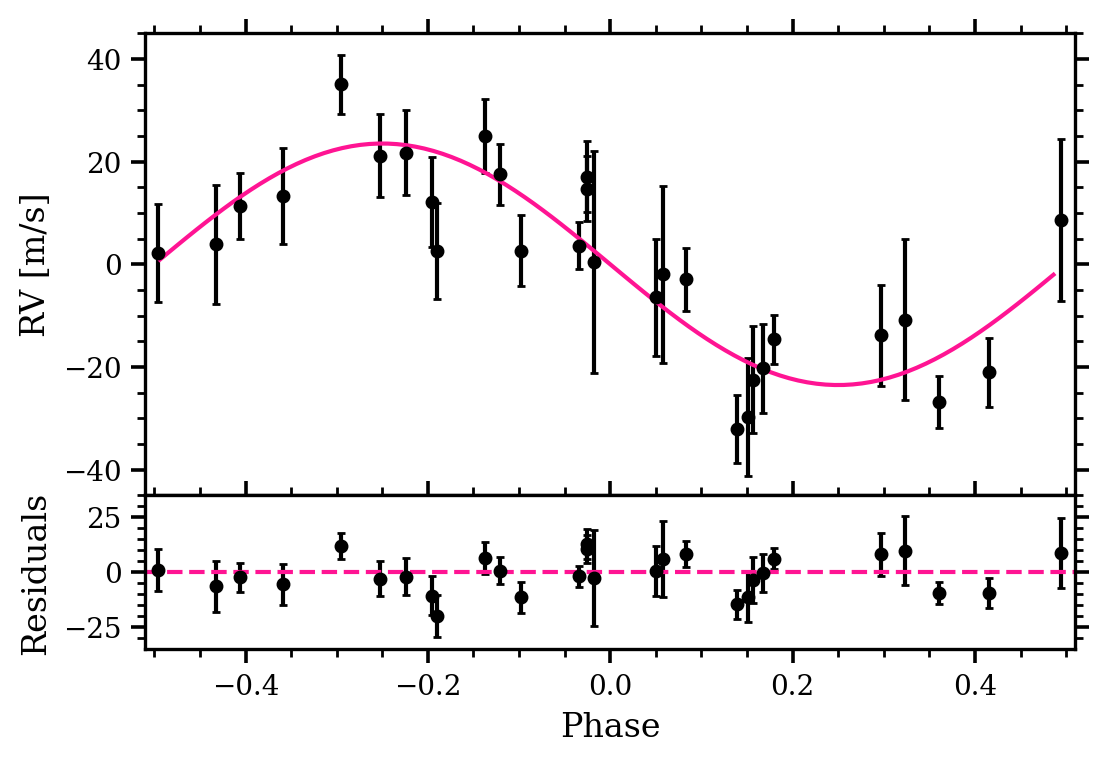}
    \caption{RVs of TOI-6038~A observed with PARAS-2 are plotted as black dots over time (left). Phase-folded RVs are shown on the right. The best-fit RV model from EXOFASTv2 is represented by the solid pink line, with the residual RVs displayed in the respective lower panels.}
    \label{fig:rv}
\end{figure*}

The best-fit Kurucz stellar atmosphere model from the SED and the best-fit MIST stellar evolutionary model are shown in Fig. \ref{fig:sed} and Fig. \ref{fig:mist}, respectively. We find that the host star is a late F-type star with parameters $M_* = 1.291^{+0.066}_{-0.060} \ \msun$, $R_* = 1.648^{+0.045}_{-0.035} \ \rsun$, $T_{\mathrm{eff}} = 6110\pm100$ K, $\log{g} = 4.118^{+0.015}_{-0.025}$, and an age of $3.65^{+0.92}_{-0.85}$ Gyr. These parameters are consistent with those derived from spectral synthesis in Sec. \ref{sec:spec_synthesis}. The best-fit transit light curve and RV models are shown in Fig. \ref{fig:tesslc} and Fig. \ref{fig:rv}, respectively. Joint modeling reveals a sub-Saturn planet with a radius $R_P = 6.41^{+0.20}_{-0.16} \ R_\oplus$ and a mass $M_P = 78.5^{+9.5}_{-9.9} \ M_\oplus$ in a circular orbit with period $P = 5.8267311^{+0.0000074}_{-0.0000068}$ days. The best-fit stellar and planetary parameters, along with their 68\% confidence intervals for the circular fit high-mass solution, are listed in Tables \ref{tab:star_table} and \ref{tab:exofast_table}, respectively. We have also listed important parameters from the low-mass solution in Table \ref{tab:exofast_table_lowmass_solution}.

\renewcommand{\arraystretch}{1.3}
\begin{table*}[t!]
\caption{Summary of EXOFASTv2 fitted and derived parameters with 68\% confidence interval for TOI-6038~A system, based on the high-mass solution (relative probability = 87\%).}
\label{tab:exofast_table}   
\centering
\begin{tabular}{llll}
\hline
\hline
\noalign{\smallskip}
Parameter&Description&Value\\
\noalign{\smallskip}
\hline
\noalign{\smallskip}
\multicolumn{2}{l}{Planetary Parameters:}\\
~~~~$P$\dotfill &Period (days)\dotfill &$5.8267311^{+0.0000074}_{-0.0000068}$\\
~~~~$R_P$\dotfill &Radius ($R_\oplus$)\dotfill &$6.41^{+0.20}_{-0.16}$\\
~~~~$M_P$\dotfill &Mass ($M_\oplus$)\dotfill &$78.5^{+9.5}_{-9.9}$\\
~~~~$T_C$\dotfill &Observed Time of conjunction (\bjdtdb)\dotfill &$2459883.05800^{+0.00072}_{-0.00070}$\\
~~~~$a$\dotfill &Semi-major axis (AU)\dotfill &$0.0690^{+0.0012}_{-0.0011}$\\
~~~~$i$\dotfill &Inclination (Degrees)\dotfill &$88.96^{+0.71}_{-0.81}$\\
~~~~$T_{\rm eq}$\dotfill &Equilibrium temp{\textdagger} (K)\dotfill &$1439^{+25}_{-24}$\\
~~~~$K$\dotfill &RV semi-amplitude (m/s)\dotfill &$23.5\pm2.8$\\
~~~~$R_P/R_*$\dotfill &Radius of planet in stellar radii \dotfill &$0.03570\pm0.00036$\\
~~~~$a/R_*$\dotfill &Semi-major axis in stellar radii \dotfill &$9.04^{+0.12}_{-0.24}$\\
~~~~$\delta$\dotfill &$\left(R_P/R_*\right)^2$ \dotfill &$0.001274^{+0.000026}_{-0.000025}$\\
~~~~$\delta_{\rm TESS}$\dotfill &Transit depth in TESS (frac)\dotfill &$0.001442^{+0.000032}_{-0.000031}$\\
~~~~$\tau$\dotfill &In/egress transit duration (days)\dotfill &$0.00746^{+0.00047}_{-0.00020}$\\
~~~~$T_{14}$\dotfill &Total transit duration (days)\dotfill &$0.2101\pm0.0014$\\
~~~~$b$\dotfill &Transit impact parameter \dotfill &$0.16^{+0.12}_{-0.11}$\\
~~~~$\rho_P$\dotfill &Density (cgs)\dotfill &$1.62^{+0.23}_{-0.24}$\\
~~~~$logg_P$\dotfill &Surface gravity (cgs)\dotfill &$3.269^{+0.053}_{-0.062}$\\
~~~~$\fave$\dotfill &Incident Flux (\fluxcgs)\dotfill &$0.975^{+0.070}_{-0.064}$\\
~~~~$M_P\sin i$\dotfill &Minimum mass ($M_\oplus$)\dotfill &$78.2^{+9.5}_{-9.9}$\\
~~~~$M_P/M_*$\dotfill &Mass ratio \dotfill &$0.000182\pm0.000022$\\
~~~~$d/R_*$\dotfill &Separation at mid transit \dotfill &$9.04^{+0.12}_{-0.24}$\\
\noalign{\smallskip}
\multicolumn{2}{l}{Wavelength Parameters (TESS):}\\
~~~~$u_{1}$\dotfill &Linear limb-darkening coeff \dotfill &$0.242^{+0.034}_{-0.035}$\\
~~~~$u_{2}$\dotfill &Quadratic limb-darkening coeff \dotfill &$0.291\pm0.035$\\
\noalign{\smallskip}
\multicolumn{2}{l}{Telescope Parameters (PARAS-2):}\\
~~~~$\gamma_{\rm rel}$\dotfill &Relative RV Offset (m/s)\dotfill &$-0.2\pm2.0$\\
~~~~$\sigma_J$\dotfill &RV Jitter (m/s)\dotfill &$6.4^{+2.2}_{-2.1}$\\
~~~~$\sigma_J^2$\dotfill &RV Jitter Variance \dotfill &$40^{+33}_{-22}$\\
\noalign{\smallskip}
\multicolumn{2}{l}{Transit Parameters (TESS):} &  Sector 18, Sector 58\\
~~~~$\sigma^{2}$\dotfill &Added Variance \dotfill & $(3.31^{+0.80}_{-0.78}, \ 2.24^{+0.49}_{-0.48})\times 10^{-8}$\\
\noalign{\smallskip}
\hline
\noalign{\smallskip}
\multicolumn{3}{l}{\footnotesize{\textbf{Notes.} {\textdagger}Assumes no albedo and perfect redistribution}}
\end{tabular}
\end{table*}

\subsection{Tidal circularization timescale} \label{tidal}

We calculate the tidal circularization timescale ($\tau_{\text{circ}}$) for TOI-6038~A~b to be $3.6 \pm 0.9$ Gyr. This is calculated using equation 2 from \cite{circ_time} and assuming a tidal quality factor of $Q_P = 10^{5}$ \citep{Petigura_2017,subjak2022}. The value of $\tau_{\text{circ}}$ is comparable to the current age of the system, which is $3.65^{+0.92}_{-0.85}$ Gyr. However, for $Q_P = 10^{5.5}$, $\tau_{\text{circ}}$ increases to $11.3 \pm 2.7$ Gyr. Therefore, given the uncertainty in the tidal quality factor, it remains unclear whether the orbit of TOI-6038~A~b is tidally circularized.


\section{Discussion} \label{sec:discussion}

\subsection{Mass-radius diagram and internal structure} \label{sec:massradius}

\begin{figure}[t!]
\centering
\includegraphics[width=0.5\textwidth]{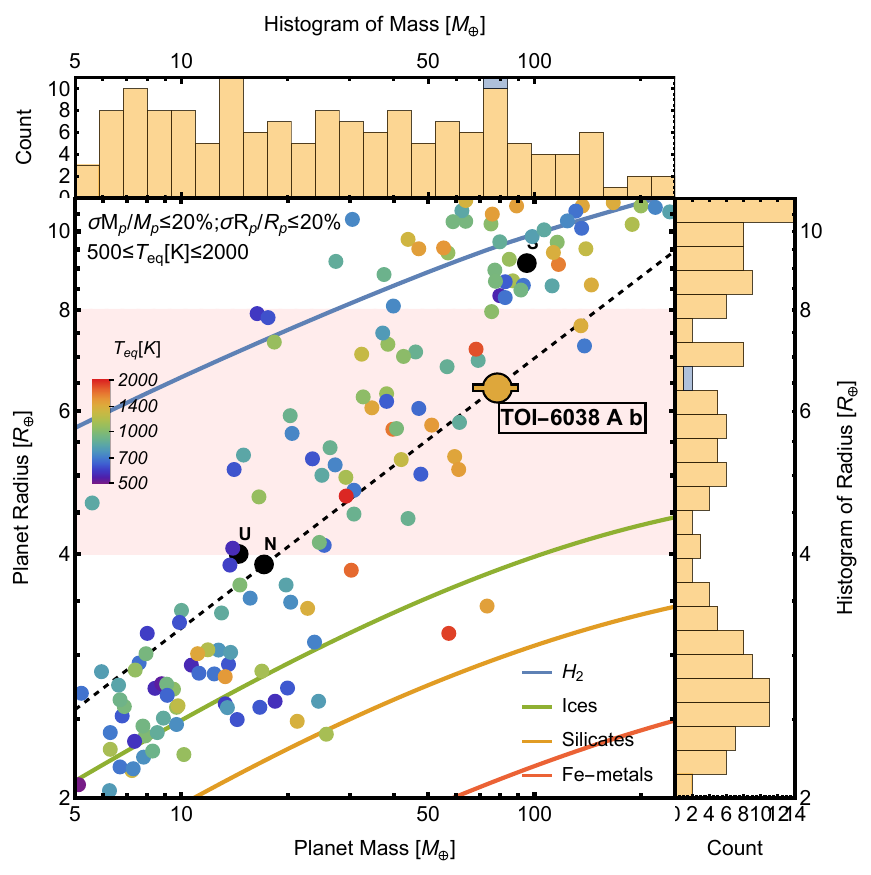}
    \caption{Mass-radius diagram of known transiting exoplanets with masses and radii constraints better than 20\%, and equilibrium temperatures ($T_{\rm eq}$) between 500 K and 2000 K. The sample is drawn from the TEPCat catalog \citep{TEPcat}. The light pink region highlights the sub-Saturn class, while the black dashed line indicates the iso-density curve for $\rho_P=1.62$. The solid curves in red, orange, green, and blue represent the mass-radius relationships for pure iron, pure silicates, pure high-pressure ices \citep{Zeng_2021}, and a pure $H_2$ core composition \citep{Becker_2014}, respectively. The Solar system planets are shown as black dots.}
    \label{fig:massradius}
\end{figure}

We plot the mass-radius (M-R) diagram of known transiting exoplanets and show the position of TOI-6038~A~b in Fig. \ref{fig:massradius}. The sample is drawn from the TEPCat catalog \citep{TEPcat} and only includes exoplanets with masses and radii constraints better than 20\%, and equilibrium temperatures ($T_{\rm eq}$) between 500 K and 2000 K. Additionally, we have overlaid theoretical M-R iso-composition curves for planets with pure iron, pure silicate, and pure high-pressure ice core compositions from \cite{Zeng_2021}, as well as a pure $H_2$ core composition at a specific entropy $S=0.3$ from \cite{Becker_2014}. TOI-6038~A~b lies well above the pure-ices composition, suggesting the presence of a significant gaseous envelope mass fraction ($f_{\rm env}$). Sub-Saturns can be modeled as two-component planets: a rocky, heavy-element core enclosed by a gaseous H/He envelope. The detailed composition does not alter $f_{\rm env}$ much, as both the heavy-element core and the low-density gaseous envelope contribute significantly to the planet’s total mass \citep{Lopez_2014,Petigura_2016}.

We used the \texttt{photoevolver} {\tt PYTHON} code \citep{Fernandez_23} to model the internal structure of TOI-6038~A~b. The code offers a range of published models to choose from. To model the core, we used the M-R empirical relation for rocky planets from \cite{Otegi20}, and for the envelope structure, we used the model from \cite{Chen_2016}, which provides the envelope radius as a function of planet mass, $f_{\rm env}$, irradiation flux, and age. The derived core and envelope parameters from the internal structure modeling are presented in Table \ref{tab:internal_structure}. Our results suggest that TOI-6038~A~b has a core mass of $\approx 58$ $M_\oplus$, composed of rocky heavy elements. The remaining mass fraction, with $f_{\rm env} \approx 0.26$ (defined as $\frac{M_{\rm P}-M_{\rm core}}{M_{\rm P}}$), consists of a low-density H/He envelope. The envelope has a radius of $3.07 \pm 0.28$ $R_\oplus$, occupying around 86\% of the planet’s total volume.

\begin{table}[t!]
    \centering
    \caption{Internal structure of TOI-6038~A~b derived using the \texttt{photoevolver}.}
    \label{tab:internal_structure}
    \begin{tabular}{lll}
         \hline
         \hline
         \noalign{\smallskip}
         Parameter & Description & Value\\
         \noalign{\smallskip}
         \hline
         \noalign{\smallskip}
         $M_{core}$ & Core mass $(M_\oplus)$ &  $58.0\pm7.8$\\
         $R_{core}$ & Core radius $(R_\oplus)$ & $3.34\pm0.20$\\
         $R_{env}$ & Envelope radius $(R_\oplus)$ & $3.07\pm0.28$\\
         $f_{env}$ & Envelope mass fraction & $0.261\pm0.036$\\
         \noalign{\smallskip}
         \hline
    \end{tabular}
\end{table}

\subsection{Formation and evolution scenarios}
\label{sec:formation}

The current orbital configuration of TOI-6038~A~b points to an interesting dynamical history. Most giant planets are expected to form beyond the ice line, where dust particles can coagulate to form massive planetesimals. The planetesimals would then accrete gas from the protoplanetary disc to form the gas giants (see \citealt{helledGiantPlanetFormation2014} for a review). After formation, a variety of mechanisms can transport the giant planets from beyond the ice line to their current close-in orbits \citep{2018ARA&A..56..175D}. One plausible scenario in the case of TOI-6038~A~b involves high eccentricity tidal migration (HEM) triggered by the secular perturbations from the stellar companion TOI-6038~B. More specifically, the von Zeipel-Lidov-Kozai (vZLK) oscillations \citep{vonzeipelLapplicationSeriesLindstedt1910,lidovEvolutionOrbitsArtificial1962,kozaiSecularPerturbationsAsteroids1962a,kinoshitaGeneralSolutionKozai2007} can be induced by the stellar companion if the stellar binary orbit and the planetary orbit are mutually misaligned. For a near circular planetary orbit, the critical mutual inclination needed for the oscillations is $\sim 40^\circ$. The critical inclination is lowered if the planetary orbit is initially eccentric. If the eccentricity is sufficiently excited by the vZLK mechanism, the planet could have been brought close to the host star, where the tidal dissipation in the planet could have circularized and shrank its orbit \citep[e.g.][]{2007ApJ...670..820W}. The characteristic timescale of the mechanism is given by $t_{vZLK} =  (a_{\rm comp}/a_{1,0})^3 (m_{\rm primary}/m_{\rm comp}) t_{P,0}$, where $t_{P,0}$ is the period of the planet. Assuming that the planet formed at around $a_{1,0} \sim 5$ AU, the secular timescale is $t_{vZLK} \sim 4 \times 10^9$ years, which is comparable to the age of the system\footnote{Here we are assuming that multiple Kozai cycles are needed to circularize the orbit.}. But, it should be noted that short-range forces like relativistic corrections, rotational and tidal deformations can suppress the eccentricity excitation caused by the vZLK mechanism and hence prevent HEM \citep{2007ApJ...670..820W}. The relative importance of these short-range forces is quantified by $\epsilon_{\rm GR}, \epsilon_{\rm rot}$ and $\epsilon_{\rm td}$, which are defined in \cite{liuSuppressionExtremeOrbital2015}. For $a_{1,0}=5$ AU, $\epsilon_{\rm GR}\sim 3$, which indicates that the vZLK oscillations are indeed suppressed by relativistic apsidal precession.  Rotational and tidal deformations are less important due to their stronger dependence on the distance between the planet and the star ($\epsilon_{\rm td}\sim 6 \times 10^{-9}$ and $\epsilon_{\rm rot} \sim 0.013$, assuming that the love number of the planet ($k_2$) is 0.37, and that the planet spin period is 10 hours). It should also be noted that the orbital parameters of the stellar binary orbit are not well constrained. In the above analysis, we used the projected separation between the stars as the semi-major axis of the stellar binary. This may underestimate the real semi-major axis of the binary. Also, if the stellar binary is significantly eccentric, the parameter space which allows HEM is enhanced \citep[e.g.][]{2016ARA&A..54..441N}. Our preliminary analysis suggests that HEM induced by the stellar companion may be unlikely. In a follow-up work we plan to study this system in detail using a set of dynamical and atmospheric simulations.

Multiple alternative mechanisms are also possible. For instance, outer massive planetary companions on co-planar eccentric orbits \citep{petrovichHotJupitersCoplanar2015} or misaligned orbits \citep{2011Natur.473..187N,2013ApJ...779..166T} can trigger HEM processes. Planet's eccentricity could also have been excited through secular chaos, which requires at least two companions \citep{2011ApJ...735..109W}. Beyond secular dynamics, planet-planet scattering may also excite planet's eccentricity \citep{2012ApJ...751..119B}. While no planetary companions have been discovered in this system so far, future observations of the system will help us either verify or rule out the above mechanisms. The planet could also have migrated throughout the protoplanetary disc soon after its formation \citep[e.g.][]{1979ApJ...233..857G,1996Natur.380..606L}. The details of disk-driven migration are highly sensitive to the disk conditions, and hence it is not clear how important this migration is for the formation of close-in giants \citep[e.g.][]{2014prpl.conf..667B,2021JGRE..12606629F}. Alternatively,  TOI-6038~A~b  could also have formed in-situ \citep{boleySITUFORMATIONGIANT2016,leeBreedingSuperEarthsBirthing2016,batyginSITUFORMATIONDYNAMICAL2016}. One of the possible ways to better constrain its formation and evolution pathway is to measure the spin-orbit misalignment of the star-planet system \citep[e.g.][]{2018haex.bookE...2T,2018Natur.553..477B,Bourrier2023}. We estimated the probability of spin-orbit misalignment for the TOI-6038~A system to be 63\%, using the analytical formula provided by \cite{Attia_2023}. If the spin of the star is aligned with the orbital angular momentum of the planet, the planet is likely to have migrated in the plane of the disk i.e., through coplanar HEM, disk-driven migration or planet-planet scattering. Meanwhile, if the planet has significant stellar obliquity, the vZLK mechanism \citep{storchChaoticDynamicsStellar2014,storchChaoticDynamicsStellar2015,andersonFormationStellarSpinorbit2016} or secular chaos \citep{2011ApJ...735..109W} might be responsible for its current close-in location.

\subsection{Contextualization in the exo-Neptunian landscape} \label{sec:neptunian_landscape}

\begin{figure}[t!]
    \centering
    \includegraphics[width=0.45\textwidth]{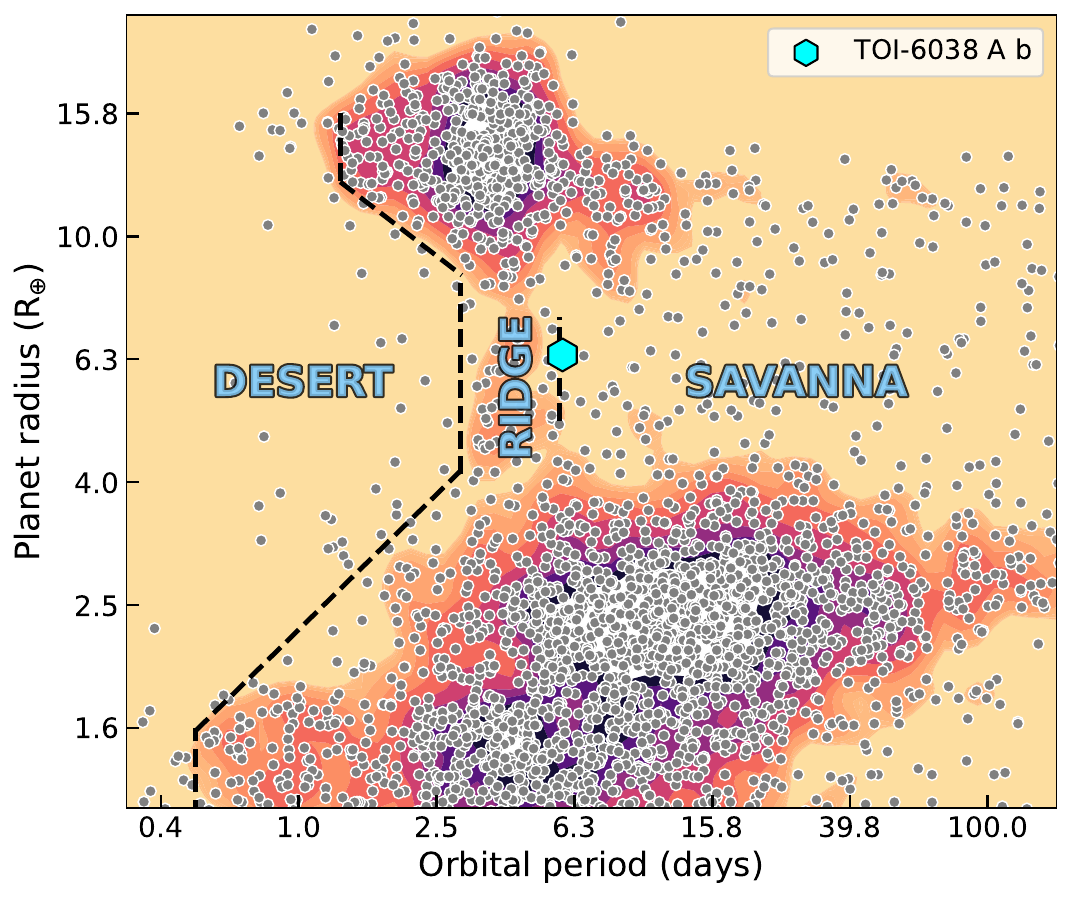}
    \caption{TOI-6038~A~b in the period-radius diagram of close-in exoplanets with radii constrained to a precision better than 20$\%$. The boundaries of the Neptunian desert, ridge, and savanna \citep{CastroGonzalez2024a} are indicated with black dashed lines. The data were collected from the NASA Exoplanet Archive \citep[][]{NASA_EXO_Archive_Akeson_2013}.}
    \label{fig:nep_des_ridge_savanna}
\end{figure}

\begin{figure}[t!]
    \centering
    \includegraphics[width=0.47\textwidth]{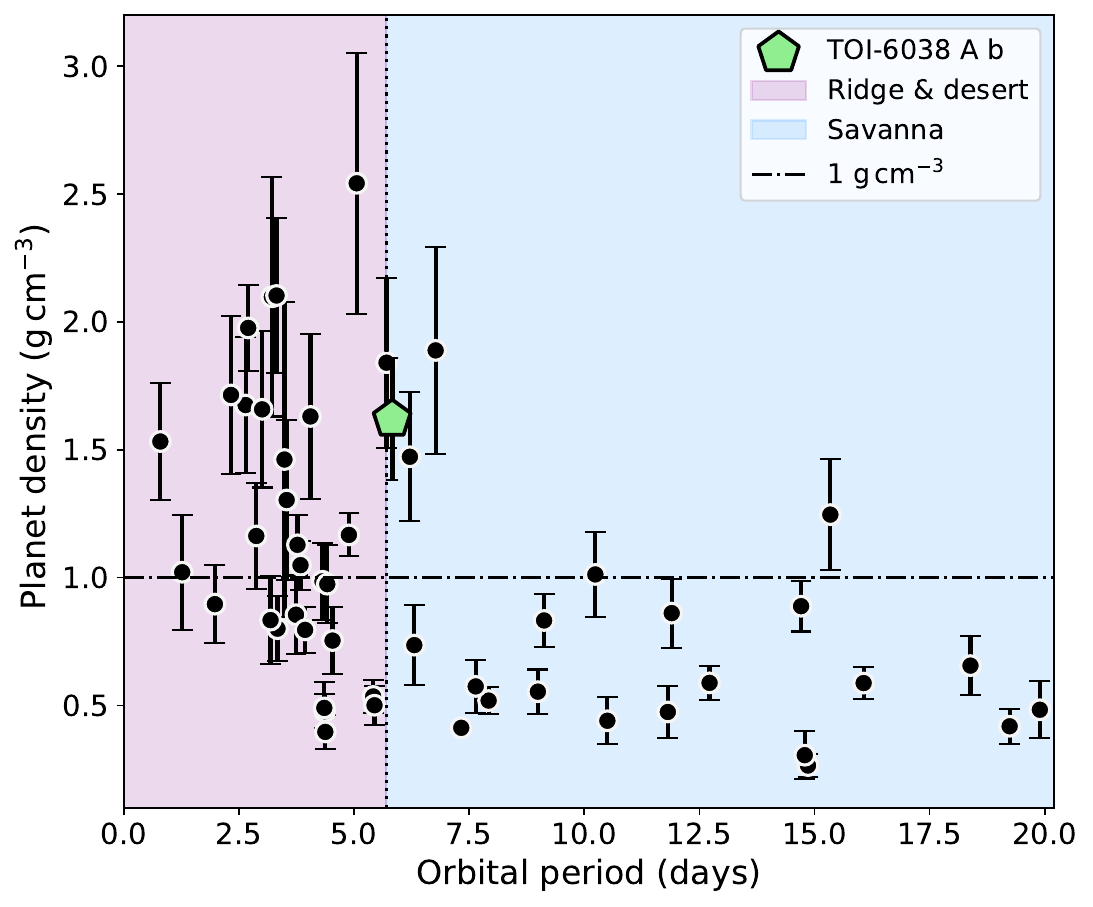}
    \caption{Density-period diagram of super-Neptunes and sub-Saturns with masses and radii constrained to a precision better than 20$\%$. Planets in the savanna show low densities, typically below $1\, \rm g\,cm^{-3}$, while planets in the ridge show densities as high as $1.5-2.0\, \rm g\,cm^{-3}$ \citep{CastroGonzalez2024b}. Data sourced from the NASA Exoplanet Archive.}
    \label{fig:dens_porb}
\end{figure}

In Fig.~\ref{fig:nep_des_ridge_savanna}, we contextualize TOI-6038~A~b in the period-radius diagram of known exoplanets with radii constrained to a precision better than 20$\%$. In recent work, \citet{CastroGonzalez2024a} conducted a planet occurrence study based on \textit{Kepler} data \citep{2010Sci...327..977B,2018ApJS..235...38T} and identified an overdensity of super-Neptunes and sub-Saturns with orbital periods between $\simeq$3.2 and $\simeq$5.7 days, which they call the Neptunian ridge. The ridge separates the Neptunian desert \citep[i.e., a sparsely populated region of intermediate planets on the shortest-period orbits;][]{2011A&A...528A...2B,2011ApJ...727L..44S,2013ApJ...763...12B,Mazeh2016} from the savanna \citep[i.e., a moderately populated region at larger orbital distances;][]{Bourrier2023}, revealing a preferred location for close-in intermediate planets. Interestingly, having an orbital period of 5.8 days, TOI-6038~A~b is located at the border between the ridge and the savanna, which makes it a target of interest to study the transition between these two regimes.

The period range of the Neptunian ridge coincides with that of the hot Jupiter pileup \citep{1999ApJ...526..890C,2003A&A...407..369U,2005ApJ...623..472G,2006ApJ...646..505B}, suggesting that similar evolutionary processes might be acting in both populations. While both disk-driven migration and HEM processes are thought to be necessary to explain the observed properties of close-in Jupiter-size planets, HEM processes are considered to play a predominant role \citep[see][for a review]{2018ARA&A..56..175D,2021JGRE..12606629F}. Interestingly, different works are showing a considerable number of eccentric and misaligned Jupiter- and Neptune-sized planets within the $\simeq$3-6 days over-densities \citep[e.g.][]{2012ApJ...757...18A,Correia_20,Bourrier2023}, which favors the hypothesis that both the pileup and the ridge could be primarily populated by HEM processes. As discussed in Sect.~\ref{sec:formation}, it is reasonable to think that the wide-orbit binary companion TOI-6038~B could have triggered HEM on TOI-6038~A~b. However, our preliminary analysis suggests that vZLK oscillations caused by the stellar companion may be suppressed by relativistic apsidal precession. A detailed analysis of the system is left to a future study. Therefore, searching for additional massive companions as well as measuring the spin-orbit angle of TOI-6038~A~b will be key to bringing tighter constraints on its overall evolution.

In addition to the planet occurrence difference between the ridge and the savanna, \citet{CastroGonzalez2024b} find a dichotomy between the densities of the planets in these two regimes. Planets in the savanna have low densities (about $\rm 0.5 \, g\,cm^{-3}$), very rarely surpassing $\rm 1 g\,cm^{-3}$, while planets in the ridge (and desert) frequently show densities as high as $\rm 1.5-2.0 \, g\,cm^{-3}$. In Fig.~\ref{fig:dens_porb}, we contextualize TOI-6038~A~b in the density-period diagram of known super-Neptunes and sub-Saturns (4$R_{\oplus}$ $<$ $R_{\rm p}$ $<$ 8$R_{\oplus}$) with masses and radii constrained to a precision better than 20$\%$. Having a high density ($\rho_{\rm P}$ = $1.62^{+0.23}_{-0.24} \rm \, g\,cm^{-3}$), and being located right at the border of the ridge, TOI-6038~A~b is consistent with the recently identified population of high-density ridge planets. The origins of this density dichotomy could be explained through atmospheric, formation, and dynamical processes. On the one hand, super-Neptunes and sub-Saturns could have all formed with low densities, below $1 \rm \, g\,cm^{-3}$, presumably with extensive H/He atmospheres, and only those receiving high levels of irradiation at short orbital distances would have been able to undergo strong enough evaporation to significantly increase the bulk density of the planet \citep[e.g.][]{2015Natur.522..459E,2018A&A...620A.147B}. On the other hand, the density dichotomy could also be explained by the existence of two different sub-populations of Neptunian planets that formed and migrated through different channels. To date, there is not enough observational evidence to distinguish between the different hypotheses. Hence, confirming and characterizing planets around the boundary between the ridge and the savanna such as TOI-6038~A~b will provide important insight into the transition between two possible populations of Neptunian planets.

\subsection{Future follow-up opportunities} \label{sec:future_followup}

The atmospheric scale height for TOI-6038~A~b is calculated to be $279 \pm 40$ km, using the equation $H_b = \frac{k T_{eq}}{\mu g_P}$ from \citet{Madhusudhan_2014}, assuming a H/He atmosphere with a mean molecular mass of $\mu = 2.3$ amu. TOI-6038~A~b has a Transmission Spectroscopic Metric (TSM) of $37 \pm 6$, which, while below the threshold of 90 set for sub-Jovian planets by \citet{Kempton_2018}, still holds promise for further investigation. Future atmospheric characterization studies will help reveal the planet’s composition, offering clues to its formation history and enhancing our understanding of this system.

The estimated RV semi-amplitude of the Rossiter-McLaughlin (RM) effect for TOI-6038~A~b is around $7\ m\ s^{-1}$ \citep{Rossiter_1924,McLaughlin_1924}, well within reach for follow-up RV transit observations to determine the spin-orbit misalignment of the star-planet system. Determining the spin-orbit angle will be instrumental in constraining the planet's migration scenarios, as discussed in Sec. \ref{sec:formation} and \ref{sec:neptunian_landscape}. Further RV measurements and transit-timing variations (TTVs) analyses from upcoming TESS data could provide valuable insights into the potential presence of additional planets in the system. Identifying planetary companions would help confirm or rule out HEM as a possible mechanism for TOI-6038~A~b’s current orbit.


\section{Summary}\label{sec:summary} 

We presented the discovery and characterization of TOI-6038~A~b, a hot sub-Saturn orbiting a metal-rich late F-type star in a wide binary system. Using a combination of RV measurements from the PARAS-2 spectrograph, transit photometry from TESS, and speckle imaging from the PRL 2.5 m telescope, we established its planetary nature and precisely determined its physical parameters. Joint modeling of the transit photometric and RV spectroscopic data yielded a planetary mass of $78.5^{+9.5}_{-9.9} \ M_\oplus$ and a radius of $6.41^{+0.20}_{-0.16} \ R_\oplus$, resulting in a bulk density of $1.62^{+0.23}_{-0.24} \rm \ g \ cm^{-3}$. Internal structure modeling indicates that TOI-6038~A~b possesses a massive core of $\approx$ 58 $M_\oplus$, composed predominantly of dense materials such as rock and iron, which account for around 74\% of the planet’s total mass. The remaining mass is composed of a modest H/He envelope, suggesting a relatively thin atmosphere.

With an orbital period of $\approx$ 5.8 days, TOI-6038~A~b lies at the boundary between the Neptunian ridge and savanna, making it a key target for studying the processes that shape these distinct populations. Our estimates suggest that secular perturbations from the wide binary companion, TOI-6038~B, are insufficient to induce complete high-eccentricity migration (HEM). However, the presence of undetected planetary companions in the system could still explain the planet's migration through HEM. Additionally, there remains a possibility that TOI-6038~A~b’s orbit was shaped by early disk-driven migration. Thus, we leave both HEM and disk-driven migration as open possibilities. Future atmospheric characterization studies and spin-orbit angle measurements, constraining joint atmospheric and dynamical evolutionary simulations \citep[e.g.][]{2021A&A...647A..40A}, will be crucial for gaining deeper insights into the formation and history of this intriguing system.


\section*{Acknowledgments}

We gratefully acknowledge the support from PRL-DOS (Department of Space, Government of India) and the Director of PRL for funding the PARAS-2 spectrograph and providing research grants for S.B. We are also thankful to the staff at the Mount Abu Observatory for their invaluable assistance during the observations. H.G.B. thanks Cristobal Petrovich, Gongjie Li, and Hagai Perets for insightful discussions. A.C.-G. is funded by the Spanish Ministry of Science through MCIN/AEI/10.13039/501100011033 grant PID2019-107061GB-C61. This work has been carried out within the framework of the NCCR PlanetS, supported by the Swiss National Science Foundation under grants 51NF40$\_$182901 and 51NF40$\_$205606, and has received funding from the European Research Council (ERC) under the European Union's Horizon 2020 research and innovation programme (project {\sc Spice Dune}, grant agreement No 947634). This research made use of the SIMBAD database and the VizieR catalog access tool, both operated by CDS in Strasbourg, France. We also utilized the Exoplanet Follow-up Observation Program (ExoFOP; DOI: 10.26134/ExoFOP5), managed by the California Institute of Technology under contract with NASA as part of the Exoplanet Exploration Program. Furthermore, this study incorporates data collected by the TESS mission, obtained from the MAST data archive at the Space Telescope Science Institute (STScI), as well as data from the TepCat \citep{TEPcat} and NASA Exoplanet Archive \citep{NASA_EXO_Archive_Akeson_2013} catalogs. This work made use of \texttt{TESS-cont} (\url{https://github.com/castro-gzlz/TESS-cont}), which also made use of \texttt{tpfplotter} \citep{tpfplot} and \texttt{TESS\_PRF} \citep{2022ascl.soft07008B}. The authors thank the anonymous referee for their valuable suggestions, which enhanced the quality of the paper.


\bibliography{ref}{}

\begin{thebibliography}{}
\expandafter\ifx\csname natexlab\endcsname\relax\def\natexlab#1{#1}\fi
\providecommand{\url}[1]{\href{#1}{#1}}
\providecommand{\dodoi}[1]{doi:~\href{http://doi.org/#1}{\nolinkurl{#1}}}
\providecommand{\doeprint}[1]{\href{http://ascl.net/#1}{\nolinkurl{http://ascl.net/#1}}}
\providecommand{\doarXiv}[1]{\href{https://arxiv.org/abs/#1}{\nolinkurl{https://arxiv.org/abs/#1}}}

\bibitem[{{Adams} \& {Laughlin}(2006)}]{circ_time}
{Adams}, F.~C., \& {Laughlin}, G. 2006, \apj, 649, 1004, \dodoi{10.1086/506145}

\bibitem[{Agol {et~al.}(2020)Agol, Luger, \& Foreman-Mackey}]{Agol_2020}
Agol, E., Luger, R., \& Foreman-Mackey, D. 2020, The Astronomical Journal, 159, 123, \dodoi{10.3847/1538-3881/ab4fee}

\bibitem[{{Akaike}(1974)}]{AIC}
{Akaike}, H. 1974, IEEE Transactions on Automatic Control, 19, 716

\bibitem[{Akeson {et~al.}(2013)Akeson, Chen, Ciardi, Crane, Good, Harbut, Jackson, Kane, Laity, Leifer, Lynn, McElroy, Papin, Plavchan, Ramírez, Rey, von Braun, Wittman, Abajian, Ali, Beichman, Beekley, Berriman, Berukoff, Bryden, Chan, Groom, Lau, Payne, Regelson, Saucedo, Schmitz, Stauffer, Wyatt, \& Zhang}]{NASA_EXO_Archive_Akeson_2013}
Akeson, R.~L., Chen, X., Ciardi, D., {et~al.} 2013, Publications of the Astronomical Society of the Pacific, 125, 989, \dodoi{10.1086/672273}

\bibitem[{{Albrecht} {et~al.}(2012){Albrecht}, {Winn}, {Johnson}, {Howard}, {Marcy}, {Butler}, {Arriagada}, {Crane}, {Shectman}, {Thompson}, {Hirano}, {Bakos}, \& {Hartman}}]{2012ApJ...757...18A}
{Albrecht}, S., {Winn}, J.~N., {Johnson}, J.~A., {et~al.} 2012, \apj, 757, 18, \dodoi{10.1088/0004-637X/757/1/18}

\bibitem[{{Aller} {et~al.}(2020){Aller}, {Lillo-Box}, {Jones}, {Miranda}, \& {Barcel{\'o} Forteza}}]{tpfplot}
{Aller}, A., {Lillo-Box}, J., {Jones}, D., {Miranda}, L.~F., \& {Barcel{\'o} Forteza}, S. 2020, \aap, 635, A128, \dodoi{10.1051/0004-6361/201937118}

\bibitem[{Anderson {et~al.}(2016)Anderson, Storch, \& Lai}]{andersonFormationStellarSpinorbit2016}
Anderson, K.~R., Storch, N.~I., \& Lai, D. 2016, 456, 3671, \dodoi{10.1093/mnras/stv2906}

\bibitem[{{Andrae} {et~al.}(2023){Andrae}, {Fouesneau, M.}, {Sordo, R.}, {Bailer-Jones, C. A. L.}, {Dharmawardena, T. E.}, {Rybizki, J.}, {De Angeli, F.}, {Lindstrøm, H. E. P.}, {Marshall, D. J.}, {Drimmel, R.}, {Korn, A.J.}, {Soubiran, C.}, {Brouillet, N.}, {Casamiquela, L.}, {Rix, H.-W.}, {Abreu Aramburu, A.}, {Álvarez, M. A.}, {Bakker, J.}, {Bellas-Velidis, I.}, {Bijaoui, A.}, {Brugaletta, E.}, {Burlacu, A.}, {Carballo, R.}, {Chaoul, L.}, {Chiavassa, A.}, {Contursi, G.}, {Cooper, W. J.}, {Creevey, O. L.}, {Dafonte, C.}, {Dapergolas, A.}, {de Laverny, P.}, {Delchambre, L.}, {Demouchy, C.}, {Edvardsson, B.}, {Frémat, Y.}, {Garabato, D.}, {García-Lario, P.}, {García-Torres, M.}, {Gavel, A.}, {Gomez, A.}, {González-Santamaría, I.}, {Hatzidimitriou, D.}, {Heiter, U.}, {Jean-Antoine Piccolo, A.}, {Kontizas, M.}, {Kordopatis, G.}, {Lanzafame, A. C.}, {Lebreton, Y.}, {Licata, E. L.}, {Livanou, E.}, {Lobel, A.}, {Lorca, A.}, {Magdaleno Romeo, A.}, {Manteiga, M.}, {Marocco, F.}, {Mary, N.}, {Nicolas, C.},
  {Ordenovic, C.}, {Pailler, F.}, {Palicio, P. A.}, {Pallas-Quintela, L.}, {Panem, C.}, {Pichon, B.}, {Poggio, E.}, {Recio-Blanco, A.}, {Riclet, F.}, {Robin, C.}, {Santoveña, R.}, {Sarro, L. M.}, {Schultheis, M. S.}, {Segol, M.}, {Silvelo, A.}, {Slezak, I.}, {Smart, R. L.}, {Süveges, M.}, {Thévenin, F.}, {Torralba Elipe, G.}, {Ulla, A.}, {Utrilla, E.}, {Vallenari, A.}, {van Dillen, E.}, {Zhao, H.}, \& {Zorec, J.}}]{GSP-phot_23}
{Andrae}, R., {Fouesneau, M.}, {Sordo, R.}, {et~al.} 2023, A\&A, 674, A27, \dodoi{10.1051/0004-6361/202243462}

\bibitem[{Attia {et~al.}(2023)Attia, Bourrier, Delisle, \& Eggenberger}]{Attia_2023}
Attia, O., Bourrier, V., Delisle, J.-B., \& Eggenberger, P. 2023, A\&A, 674, A120, \dodoi{10.1051/0004-6361/202245237}

\bibitem[{{Attia} {et~al.}(2021){Attia}, {Bourrier}, {Eggenberger}, {Mordasini}, {Beust}, \& {Ehrenreich}}]{2021A&A...647A..40A}
{Attia}, O., {Bourrier}, V., {Eggenberger}, P., {et~al.} 2021, \aap, 647, A40, \dodoi{10.1051/0004-6361/202039452}

\bibitem[{Baliwal {et~al.}(2024)Baliwal, Sharma, Chakraborty, Khandelwal, Nikitha, Safonov, Strakhov, Montalto, Eastman, Latham, Bieryla, Prasad, Bharadwaj, Lad, Das, \& Nayak}]{toi6651}
Baliwal, S., Sharma, R., Chakraborty, A., {et~al.} 2024, A\&A, 691, A12, \dodoi{10.1051/0004-6361/202450934}

\bibitem[{{Baranne} {et~al.}(1996){Baranne}, {Queloz}, {Mayor}, {Adrianzyk}, {Knispel}, {Kohler}, {Lacroix}, {Meunier}, {Rimbaud}, \& {Vin}}]{1996A&AS..119..373B}
{Baranne}, A., {Queloz}, D., {Mayor}, M., {et~al.} 1996, \aaps, 119, 373

\bibitem[{Barragán {et~al.}(2021)Barragán, Aigrain, Rajpaul, \& Zicher}]{citlalicue}
Barragán, O., Aigrain, S., Rajpaul, V.~M., \& Zicher, N. 2021, Monthly Notices of the Royal Astronomical Society, 509, 866, \dodoi{10.1093/mnras/stab2889}

\bibitem[{Baruteau {et~al.}(2014)Baruteau, Crida, Paardekooper, Masset, Guilet, Bitsch, Nelson, Kley, \& Papaloizou}]{2014prpl.conf..667B}
Baruteau, C., Crida, A., Paardekooper, S.~J., {et~al.} 2014, in Protostars and Planets {{VI}}, ed. H.~Beuther, R.~S. Klessen, C.~P. Dullemond, \& T.~Henning, 667--689, \dodoi{10.2458/azu_uapress_9780816531240-ch029}

\bibitem[{Batygin {et~al.}(2016)Batygin, Bodenheimer, \& Laughlin}]{batyginSITUFORMATIONDYNAMICAL2016}
Batygin, K., Bodenheimer, P.~H., \& Laughlin, G.~P. 2016, The Astrophysical Journal, 829, 114, \dodoi{10.3847/0004-637X/829/2/114}

\bibitem[{Beaug{\'e} \& Nesvorn{\'y}(2012)}]{2012ApJ...751..119B}
Beaug{\'e}, C., \& Nesvorn{\'y}, D. 2012, 751, 119, \dodoi{10.1088/0004-637X/751/2/119}

\bibitem[{{Beaug{\'e}} \& {Nesvorn{\'y}}(2013)}]{2013ApJ...763...12B}
{Beaug{\'e}}, C., \& {Nesvorn{\'y}}, D. 2013, \apj, 763, 12, \dodoi{10.1088/0004-637X/763/1/12}

\bibitem[{Becker {et~al.}(2014)Becker, Lorenzen, Fortney, Nettelmann, Schöttler, \& Redmer}]{Becker_2014}
Becker, A., Lorenzen, W., Fortney, J.~J., {et~al.} 2014, The Astrophysical Journal Supplement Series, 215, 21, \dodoi{10.1088/0067-0049/215/2/21}

\bibitem[{{Bell} \& {Higgins}(2022)}]{2022ascl.soft07008B}
{Bell}, K.~J., \& {Higgins}, M.~E. 2022, {TESS\_PRF: Display the TESS pixel response function}, Astrophysics Source Code Library, record ascl:2207.008

\bibitem[{{Ben{\'\i}tez-Llambay} {et~al.}(2011){Ben{\'\i}tez-Llambay}, {Masset}, \& {Beaug{\'e}}}]{2011A&A...528A...2B}
{Ben{\'\i}tez-Llambay}, P., {Masset}, F., \& {Beaug{\'e}}, C. 2011, \aap, 528, A2, \dodoi{10.1051/0004-6361/201015774}

\bibitem[{{Bensby} {et~al.}(2014){Bensby}, {Feltzing}, \& {Oey}}]{Bensby2014}
{Bensby}, T., {Feltzing}, S., \& {Oey}, M.~S. 2014, \aap, 562, A71, \dodoi{10.1051/0004-6361/201322631}

\bibitem[{{Boisse} {et~al.}(2012){Boisse}, {Bonfils}, \& {Santos}}]{2012A&A...545A.109B}
{Boisse}, I., {Bonfils}, X., \& {Santos}, N.~C. 2012, \aap, 545, A109, \dodoi{10.1051/0004-6361/201219115}

\bibitem[{Boley {et~al.}(2016)Boley, Contreras, \& Gladman}]{boleySITUFORMATIONGIANT2016}
Boley, A.~C., Contreras, A. P.~G., \& Gladman, B. 2016, The Astrophysical Journal Letters, 817, L17, \dodoi{10.3847/2041-8205/817/2/L17}

\bibitem[{{Borucki} {et~al.}(2010){Borucki}, {Koch}, {Basri}, {Batalha}, {Brown}, {Caldwell}, {Caldwell}, {Christensen-Dalsgaard}, {Cochran}, {DeVore}, {Dunham}, {Dupree}, {Gautier}, {Geary}, {Gilliland}, {Gould}, {Howell}, {Jenkins}, {Kondo}, {Latham}, {Marcy}, {Meibom}, {Kjeldsen}, {Lissauer}, {Monet}, {Morrison}, {Sasselov}, {Tarter}, {Boss}, {Brownlee}, {Owen}, {Buzasi}, {Charbonneau}, {Doyle}, {Fortney}, {Ford}, {Holman}, {Seager}, {Steffen}, {Welsh}, {Rowe}, {Anderson}, {Buchhave}, {Ciardi}, {Walkowicz}, {Sherry}, {Horch}, {Isaacson}, {Everett}, {Fischer}, {Torres}, {Johnson}, {Endl}, {MacQueen}, {Bryson}, {Dotson}, {Haas}, {Kolodziejczak}, {Van Cleve}, {Chandrasekaran}, {Twicken}, {Quintana}, {Clarke}, {Allen}, {Li}, {Wu}, {Tenenbaum}, {Verner}, {Bruhweiler}, {Barnes}, \& {Prsa}}]{2010Sci...327..977B}
{Borucki}, W.~J., {Koch}, D., {Basri}, G., {et~al.} 2010, Science, 327, 977, \dodoi{10.1126/science.1185402}

\bibitem[{{Bourrier} {et~al.}(2018{\natexlab{a}}){Bourrier}, {Lovis}, {Beust}, {Ehrenreich}, {Henry}, {Astudillo-Defru}, {Allart}, {Bonfils}, {S{\'e}gransan}, {Delfosse}, {Cegla}, {Wyttenbach}, {Heng}, {Lavie}, \& {Pepe}}]{2018Natur.553..477B}
{Bourrier}, V., {Lovis}, C., {Beust}, H., {et~al.} 2018{\natexlab{a}}, \nat, 553, 477, \dodoi{10.1038/nature24677}

\bibitem[{{Bourrier} {et~al.}(2018{\natexlab{b}}){Bourrier}, {Lecavelier des Etangs}, {Ehrenreich}, {Sanz-Forcada}, {Allart}, {Ballester}, {Buchhave}, {Cohen}, {Deming}, {Evans}, {Garc{\'\i}a Mu{\~n}oz}, {Henry}, {Kataria}, {Lavvas}, {Lewis}, {L{\'o}pez-Morales}, {Marley}, {Sing}, \& {Wakeford}}]{2018A&A...620A.147B}
{Bourrier}, V., {Lecavelier des Etangs}, A., {Ehrenreich}, D., {et~al.} 2018{\natexlab{b}}, \aap, 620, A147, \dodoi{10.1051/0004-6361/201833675}

\bibitem[{{Bourrier} {et~al.}(2023){Bourrier}, {Attia}, {Mallonn}, {Marret}, {Lendl}, {Konig}, {Krenn}, {Cretignier}, {Allart}, {Henry}, {Bryant}, {Leleu}, {Nielsen}, {Hebrard}, {Hara}, {Ehrenreich}, {Seidel}, {dos Santos}, {Lovis}, {Bayliss}, {Cegla}, {Dumusque}, {Boisse}, {Boucher}, {Bouchy}, {Pepe}, {Lavie}, {Rey Cerda}, {S{\'e}gransan}, {Udry}, \& {Vrignaud}}]{Bourrier2023}
{Bourrier}, V., {Attia}, O., {Mallonn}, M., {et~al.} 2023, \aap, 669, A63, \dodoi{10.1051/0004-6361/202245004}

\bibitem[{{Buchhave} {et~al.}(2010){Buchhave}, {Bakos}, {Hartman}, {Torres}, {Kov{\'a}cs}, {Latham}, {Noyes}, {Esquerdo}, {Everett}, {Howard}, {Marcy}, {Fischer}, {Johnson}, {Andersen}, {F{\H{u}}r{\'e}sz}, {Perumpilly}, {Sasselov}, {Stefanik}, {B{\'e}ky}, {L{\'a}z{\'a}r}, {Papp}, \& {S{\'a}ri}}]{buchhave2010}
{Buchhave}, L.~A., {Bakos}, G.~{\'A}., {Hartman}, J.~D., {et~al.} 2010, \apj, 720, 1118, \dodoi{10.1088/0004-637X/720/2/1118}

\bibitem[{Buchhave {et~al.}(2012)Buchhave, Latham, Johansen, Bizzarro, Torres, Rowe, Batalha, Borucki, Brugamyer, Caldwell, Bryson, Ciardi, Cochran, Endl, Esquerdo, Ford, Geary, Gilliland, Hansen, Isaacson, Laird, Lucas, Marcy, Morse, Robertson, Shporer, Stefanik, Still, \& Quinn}]{buchhave2012}
Buchhave, L.~A., Latham, D., Johansen, A., {et~al.} 2012, Nature, 486, 375

\bibitem[{{Buchhave} {et~al.}(2014){Buchhave}, {Bizzarro}, {Latham}, {Sasselov}, {Cochran}, {Endl}, {Isaacson}, {Juncher}, \& {Marcy}}]{buchhave2014}
{Buchhave}, L.~A., {Bizzarro}, M., {Latham}, D.~W., {et~al.} 2014, \nat, 509, 593, \dodoi{10.1038/nature13254}

\bibitem[{{Burnham} \& {Anderson}(2002)}]{BIC}
{Burnham}, K.~P., \& {Anderson}, D.~R. 2002, {Model Selection and Multimodel Inference}, 2nd edn. ({New York}: {Springer-Verlag}), \dodoi{10.1007/b97636}

\bibitem[{{Butler} {et~al.}(2006){Butler}, {Wright}, {Marcy}, {Fischer}, {Vogt}, {Tinney}, {Jones}, {Carter}, {Johnson}, {McCarthy}, \& {Penny}}]{2006ApJ...646..505B}
{Butler}, R.~P., {Wright}, J.~T., {Marcy}, G.~W., {et~al.} 2006, \apj, 646, 505, \dodoi{10.1086/504701}

\bibitem[{Caldwell {et~al.}(2020)Caldwell, Tenenbaum, Twicken, Jenkins, Ting, Smith, Hedges, Fausnaugh, Rose, \& Burke}]{tess-spoc}
Caldwell, D.~A., Tenenbaum, P., Twicken, J.~D., {et~al.} 2020, Research Notes of the AAS, 4, 201, \dodoi{10.3847/2515-5172/abc9b3}

\bibitem[{Carmichael {et~al.}(2021)Carmichael, Quinn, Zhou, Grieves, Irwin, Stassun, Vanderburg, Winn, Bouchy, Brasseur, Briceño, Caldwell, Charbonneau, Collins, Colon, Eastman, Fausnaugh, Fong, Fűrész, Huang, Jenkins, Kielkopf, Latham, Law, Lund, Mann, Ricker, Rodriguez, Schwarz, Shporer, Tenenbaum, Wood, \& Ziegler}]{Carmichael_2021}
Carmichael, T.~W., Quinn, S.~N., Zhou, G., {et~al.} 2021, The Astronomical Journal, 161, 97, \dodoi{10.3847/1538-3881/abd4e1}

\bibitem[{{Castro-Gonz{\'a}lez} {et~al.}(2024{\natexlab{a}}){Castro-Gonz{\'a}lez}, {Bourrier}, {Lillo-Box}, {Delisle}, {Armstrong}, {Barrado}, \& {Correia}}]{CastroGonzalez2024a}
{Castro-Gonz{\'a}lez}, A., {Bourrier}, V., {Lillo-Box}, J., {et~al.} 2024{\natexlab{a}}, \aap, 689, A250, \dodoi{10.1051/0004-6361/202450957}

\bibitem[{{Castro-Gonz{\'a}lez} {et~al.}(2022){Castro-Gonz{\'a}lez}, {D{\'\i}ez Alonso}, {Men{\'e}ndez Blanco}, {Livingston}, {de Leon}, {Lillo-Box}, {Korth}, {Fern{\'a}ndez Men{\'e}ndez}, {Recio}, {Izquierdo-Ruiz}, {Coya Lozano}, {Garc{\'\i}a de la Cuesta}, {G{\'o}mez Hern{\'a}ndez}, {Vidal Blanco}, {Hevia D{\'\i}az}, {Pardo Silva}, {P{\'e}rez Acevedo}, {Polancos Ruiz}, {Padilla Tijer{\'\i}n}, {V{\'a}zquez Garc{\'\i}a}, {Su{\'a}rez G{\'o}mez}, {Garc{\'\i}a Riesgo}, {Gonz{\'a}lez Guti{\'e}rrez}, {Bonavera}, {Gonz{\'a}lez-Nuevo}, {Rodr{\'\i}guez Pereira}, {S{\'a}nchez Lasheras}, {S{\'a}nchez Rodr{\'\i}guez}, {Mu{\~n}iz}, {Santos Rodr{\'\i}guez}, \& {de Cos Juez}}]{2022MNRAS.509.1075C}
{Castro-Gonz{\'a}lez}, A., {D{\'\i}ez Alonso}, E., {Men{\'e}ndez Blanco}, J., {et~al.} 2022, \mnras, 509, 1075, \dodoi{10.1093/mnras/stab2669}

\bibitem[{{Castro-Gonz{\'a}lez} {et~al.}(2023){Castro-Gonz{\'a}lez}, {Demangeon}, {Lillo-Box}, {Lovis}, {Lavie}, {Adibekyan}, {Acu{\~n}a}, {Deleuil}, {Aguichine}, {Zapatero Osorio}, {Tabernero}, {Davoult}, {Alibert}, {Santos}, {Sousa}, {Antoniadis-Karnavas}, {Borsa}, {Winn}, {Allende Prieto}, {Figueira}, {Jenkins}, {Sozzetti}, {Damasso}, {Silva}, {Astudillo-Defru}, {Barros}, {Bonfils}, {Cristiani}, {Di Marcantonio}, {Gonz{\'a}lez Hern{\'a}ndez}, {Curto}, {Martins}, {Nunes}, {Palle}, {Pepe}, {Seager}, \& {Su{\'a}rez Mascare{\~n}o}}]{2023A&A...675A..52C}
{Castro-Gonz{\'a}lez}, A., {Demangeon}, O.~D.~S., {Lillo-Box}, J., {et~al.} 2023, \aap, 675, A52, \dodoi{10.1051/0004-6361/202346550}

\bibitem[{{Castro-Gonz{\'a}lez} {et~al.}(2024{\natexlab{b}}){Castro-Gonz{\'a}lez}, {Lillo-Box}, {Armstrong}, {Acu{\~n}a}, {Aguichine}, {Bourrier}, {Gandhi}, {Sousa}, {Delgado-Mena}, {Moya}, {Adibekyan}, {Correia}, {Barrado}, {Damasso}, {Winn}, {Santos}, {Barkaoui}, {Barros}, {Benkhaldoun}, {Bouchy}, {Brice{\~n}o}, {Caldwell}, {Collins}, {Essack}, {Ghachoui}, {Gillon}, {Hounsell}, {Jehin}, {Jenkins}, {Keniger}, {Law}, {Mann}, {Nielsen}, {Pozuelos}, {Schanche}, {Seager}, {Tan}, {Timmermans}, {Villase{\~n}or}, {Watkins}, \& {Ziegler}}]{CastroGonzalez2024b}
{Castro-Gonz{\'a}lez}, A., {Lillo-Box}, J., {Armstrong}, D.~J., {et~al.} 2024{\natexlab{b}}, \aap, 691, A233, \dodoi{10.1051/0004-6361/202451656}

\bibitem[{Chakraborty {et~al.}(2018)Chakraborty, Thapa, Kumar, Neelam, Sharma, \& Roy}]{paras2_design}
Chakraborty, A., Thapa, N., Kumar, K., {et~al.} 2018, in Ground-based and Airborne Instrumentation for Astronomy VII, ed. C.~J. Evans, L.~Simard, \& H.~Takami, Vol. 10702, International Society for Optics and Photonics (SPIE), 107026G, \dodoi{10.1117/12.2313055}

\bibitem[{Chakraborty {et~al.}(2024)Chakraborty, Bharadwaj, Prasad, Sharma, Lad, Nayak, Jithendran, Joshi, Mishra, \& Ahmed}]{prl2.5m_and_paras2}
Chakraborty, A., Bharadwaj, K.~K., Prasad, N. J. S. S.~V., {et~al.} 2024, The PRL 2.5m Telescope and its First Light Instruments: FOC \& PARAS-2.
\newblock \doarXiv{2401.07715}

\bibitem[{Chen \& Rogers(2016)}]{Chen_2016}
Chen, H., \& Rogers, L.~A. 2016, The Astrophysical Journal, 831, 180, \dodoi{10.3847/0004-637X/831/2/180}

\bibitem[{{Choi} {et~al.}(2016){Choi}, {Dotter}, {Conroy}, {Cantiello}, {Paxton}, \& {Johnson}}]{mist_choi}
{Choi}, J., {Dotter}, A., {Conroy}, C., {et~al.} 2016, \apj, 823, 102, \dodoi{10.3847/0004-637X/823/2/102}

\bibitem[{{Claret}(2017)}]{Claret_tess}
{Claret}, A. 2017, \aap, 600, A30, \dodoi{10.1051/0004-6361/201629705}

\bibitem[{{Claret} \& {Bloemen}(2011)}]{Claret}
{Claret}, A., \& {Bloemen}, S. 2011, \aap, 529, A75, \dodoi{10.1051/0004-6361/201116451}

\bibitem[{Correia {et~al.}(2020)Correia, Bourrier, \& Delisle}]{Correia_20}
Correia, A. C.~M., Bourrier, V., \& Delisle, J.-B. 2020, A\&A, 635, A37, \dodoi{10.1051/0004-6361/201936967}

\bibitem[{{Correia} {et~al.}(2011){Correia}, {Laskar}, {Farago}, \& {Bou{\'e}}}]{2011CeMDA.111..105C}
{Correia}, A. C.~M., {Laskar}, J., {Farago}, F., \& {Bou{\'e}}, G. 2011, Celestial Mechanics and Dynamical Astronomy, 111, 105, \dodoi{10.1007/s10569-011-9368-9}

\bibitem[{{Cumming} {et~al.}(1999){Cumming}, {Marcy}, \& {Butler}}]{1999ApJ...526..890C}
{Cumming}, A., {Marcy}, G.~W., \& {Butler}, R.~P. 1999, \apj, 526, 890, \dodoi{10.1086/308020}

\bibitem[{{Cutri} {et~al.}(2003){Cutri}, {Skrutskie}, {van Dyk}, {Beichman}, {Carpenter}, {Chester}, {Cambresy}, {Evans}, {Fowler}, {Gizis}, {Howard}, {Huchra}, {Jarrett}, {Kopan}, {Kirkpatrick}, {Light}, {Marsh}, {McCallon}, {Schneider}, {Stiening}, {Sykes}, {Weinberg}, {Wheaton}, {Wheelock}, \& {Zacarias}}]{JHK}
{Cutri}, R.~M., {Skrutskie}, M.~F., {van Dyk}, S., {et~al.} 2003, VizieR Online Data Catalog, II/246

\bibitem[{{Cutri} {et~al.}(2021){Cutri}, {Wright}, {Conrow}, {Fowler}, {Eisenhardt}, {Grillmair}, {Kirkpatrick}, {Masci}, {McCallon}, {Wheelock}, {Fajardo-Acosta}, {Yan}, {Benford}, {Harbut}, {Jarrett}, {Lake}, {Leisawitz}, {Ressler}, {Stanford}, {Tsai}, {Liu}, {Helou}, {Mainzer}, {Gettngs}, {Gonzalez}, {Hoffman}, {Marsh}, {Padgett}, {Skrutskie}, {Beck}, {Papin}, \& {Wittman}}]{ALLWISE}
{Cutri}, R.~M., {Wright}, E.~L., {Conrow}, T., {et~al.} 2021, VizieR Online Data Catalog, II/328

\bibitem[{{Damasso} {et~al.}(2019){Damasso}, {Pinamonti}, {Scandariato}, \& {Sozzetti}}]{2019MNRAS.489.2555D}
{Damasso}, M., {Pinamonti}, M., {Scandariato}, G., \& {Sozzetti}, A. 2019, \mnras, 489, 2555, \dodoi{10.1093/mnras/stz2216}

\bibitem[{{Damasso} {et~al.}(2023){Damasso}, {Rodrigues}, {Castro-Gonz{\'a}lez}, {Lavie}, {Davoult}, {Zapatero Osorio}, {Dou}, {Sousa}, {Owen}, {Sossi}, {Adibekyan}, {Osborn}, {Leinhardt}, {Alibert}, {Lovis}, {Delgado Mena}, {Sozzetti}, {Barros}, {Bossini}, {Ziegler}, {Ciardi}, {Matthews}, {Carter}, {Lillo-Box}, {Su{\'a}rez Mascare{\~n}o}, {Cristiani}, {Pepe}, {Rebolo}, {Santos}, {Allende Prieto}, {Benatti}, {Bouchy}, {Brice{\~n}o}, {Di Marcantonio}, {D'Odorico}, {Dumusque}, {Egger}, {Ehrenreich}, {Faria}, {Figueira}, {G{\'e}nova Santos}, {Gonzales}, {Gonz{\'a}lez Hern{\'a}ndez}, {Law}, {Lo Curto}, {Mann}, {Martins}, {Mehner}, {Micela}, {Molaro}, {Nunes}, {Palle}, {Poretti}, {Schlieder}, \& {Udry}}]{2023A&A...679A..33D}
{Damasso}, M., {Rodrigues}, J., {Castro-Gonz{\'a}lez}, A., {et~al.} 2023, \aap, 679, A33, \dodoi{10.1051/0004-6361/202347240}

\bibitem[{Dawson \& Johnson(2018)}]{2018ARA&A..56..175D}
Dawson, R.~I., \& Johnson, J.~A. 2018, 56, 175, \dodoi{10.1146/annurev-astro-081817-051853}

\bibitem[{{de Leon} {et~al.}(2021){de Leon}, {Livingston}, {Endl}, {Cochran}, {Hirano}, {Garc{\'\i}a}, {Mathur}, {Lam}, {Korth}, {Trani}, {Dai}, {D{\'\i}ez Alonso}, {Castro-Gonz{\'a}lez}, {Fridlund}, {Fukui}, {Gandolfi}, {Kabath}, {Kuzuhara}, {Luque}, {Savel}, {Gill}, {Dressing}, {Giacalone}, {Narita}, {Palle}, {Van Eylen}, \& {Tamura}}]{2021MNRAS.508..195D}
{de Leon}, J.~P., {Livingston}, J., {Endl}, M., {et~al.} 2021, \mnras, 508, 195, \dodoi{10.1093/mnras/stab2305}

\bibitem[{{Dotter}(2016)}]{mist_dotter}
{Dotter}, A. 2016, \apjs, 222, 8, \dodoi{10.3847/0067-0049/222/1/8}

\bibitem[{{Dumusque} {et~al.}(2014){Dumusque}, {Boisse}, \& {Santos}}]{2014ApJ...796..132D}
{Dumusque}, X., {Boisse}, I., \& {Santos}, N.~C. 2014, \apj, 796, 132, \dodoi{10.1088/0004-637X/796/2/132}

\bibitem[{{Eastman} {et~al.}(2019){Eastman}, {Rodriguez}, {Agol}, {Stassun}, {Beatty}, {Vanderburg}, {Gaudi}, {Collins}, \& {Luger}}]{exofast}
{Eastman}, J.~D., {Rodriguez}, J.~E., {Agol}, E., {et~al.} 2019, arXiv e-prints, arXiv:1907.09480.
\newblock \doarXiv{1907.09480}

\bibitem[{{Ehrenreich} {et~al.}(2015){Ehrenreich}, {Bourrier}, {Wheatley}, {Lecavelier des Etangs}, {H{\'e}brard}, {Udry}, {Bonfils}, {Delfosse}, {D{\'e}sert}, {Sing}, \& {Vidal-Madjar}}]{2015Natur.522..459E}
{Ehrenreich}, D., {Bourrier}, V., {Wheatley}, P.~J., {et~al.} 2015, \nat, 522, 459, \dodoi{10.1038/nature14501}

\bibitem[{Fernández {et~al.}(2023)Fernández, Wheatley, \& King}]{Fernandez_23}
Fernández, J.~F., Wheatley, P.~J., \& King, G.~W. 2023, Monthly Notices of the Royal Astronomical Society, 522, 4251, \dodoi{10.1093/mnras/stad1257}

\bibitem[{Fischer {et~al.}(2016)Fischer, Anglada-Escude, Arriagada, Baluev, Bean, Bouchy, Buchhave, Carroll, Chakraborty, Crepp, Dawson, Diddams, Dumusque, Eastman, Endl, Figueira, Ford, Foreman-Mackey, Fournier, Fűrész, Gaudi, Gregory, Grundahl, Hatzes, Hébrard, Herrero, Hogg, Howard, Johnson, Jorden, Jurgenson, Latham, Laughlin, Loredo, Lovis, Mahadevan, McCracken, Pepe, Perez, Phillips, Plavchan, Prato, Quirrenbach, Reiners, Robertson, Santos, Sawyer, Segransan, Sozzetti, Steinmetz, Szentgyorgyi, Udry, Valenti, Wang, Wittenmyer, \& Wright}]{Fischer_2016}
Fischer, D.~A., Anglada-Escude, G., Arriagada, P., {et~al.} 2016, Publications of the Astronomical Society of the Pacific, 128, 066001, \dodoi{10.1088/1538-3873/128/964/066001}

\bibitem[{{Ford}(2006)}]{Gelman_rubin2006}
{Ford}, E.~B. 2006, \apj, 642, 505, \dodoi{10.1086/500802}

\bibitem[{Ford \& Rasio(2008)}]{Ford_2008}
Ford, E.~B., \& Rasio, F.~A. 2008, The Astrophysical Journal, 686, 621, \dodoi{10.1086/590926}

\bibitem[{{Fortney} {et~al.}(2021){Fortney}, {Dawson}, \& {Komacek}}]{2021JGRE..12606629F}
{Fortney}, J.~J., {Dawson}, R.~I., \& {Komacek}, T.~D. 2021, Journal of Geophysical Research (Planets), 126, e06629, \dodoi{10.1029/2020JE006629}

\bibitem[{{Forveille} {et~al.}(2009){Forveille}, {Bonfils}, {Delfosse}, {Gillon}, {Udry}, {Bouchy}, {Lovis}, {Mayor}, {Pepe}, {Perrier}, {Queloz}, {Santos}, \& {Bertaux}}]{2009A&A...493..645F}
{Forveille}, T., {Bonfils}, X., {Delfosse}, X., {et~al.} 2009, \aap, 493, 645, \dodoi{10.1051/0004-6361:200810557}

\bibitem[{{Gagn{\'e}} {et~al.}(2018){Gagn{\'e}}, {Mamajek}, {Malo}, {Riedel}, {Rodriguez}, {Lafreni{\`e}re}, {Faherty}, {Roy-Loubier}, {Pueyo}, {Robin}, \& {Doyon}}]{Gagne2018}
{Gagn{\'e}}, J., {Mamajek}, E.~E., {Malo}, L., {et~al.} 2018, \apj, 856, 23, \dodoi{10.3847/1538-4357/aaae09}

\bibitem[{{Gaia Collaboration} {et~al.}(2021){Gaia Collaboration}, {Brown}, {Vallenari}, {Prusti}, {de Bruijne}, {Babusiaux}, {Biermann}, {Creevey}, {Evans}, {Eyer}, {Hutton}, {Jansen}, {Jordi}, {Klioner}, {Lammers}, {Lindegren}, {Luri}, {Mignard}, {Panem}, {Pourbaix}, {Randich}, {Sartoretti}, {Soubiran}, {Walton}, {Arenou}, {Bailer-Jones}, {Bastian}, {Cropper}, {Drimmel}, {Katz}, {Lattanzi}, {van Leeuwen}, {Bakker}, {Cacciari}, {Casta{\~n}eda}, {De Angeli}, {Ducourant}, {Fabricius}, {Fouesneau}, {Fr{\'e}mat}, {Guerra}, {Guerrier}, {Guiraud}, {Jean-Antoine Piccolo}, {Masana}, {Messineo}, {Mowlavi}, {Nicolas}, {Nienartowicz}, {Pailler}, {Panuzzo}, {Riclet}, {Roux}, {Seabroke}, {Sordo}, {Tanga}, {Th{\'e}venin}, {Gracia-Abril}, {Portell}, {Teyssier}, {Altmann}, {Andrae}, {Bellas-Velidis}, {Benson}, {Berthier}, {Blomme}, {Brugaletta}, {Burgess}, {Busso}, {Carry}, {Cellino}, {Cheek}, {Clementini}, {Damerdji}, {Davidson}, {Delchambre}, {Dell'Oro}, {Fern{\'a}ndez-Hern{\'a}ndez}, {Galluccio}, {Garc{\'\i}a-Lario},
  {Garcia-Reinaldos}, {Gonz{\'a}lez-N{\'u}{\~n}ez}, {Gosset}, {Haigron}, {Halbwachs}, {Hambly}, {Harrison}, {Hatzidimitriou}, {Heiter}, {Hern{\'a}ndez}, {Hestroffer}, {Hodgkin}, {Holl}, {Jan{\ss}en}, {Jevardat de Fombelle}, {Jordan}, {Krone-Martins}, {Lanzafame}, {L{\"o}ffler}, {Lorca}, {Manteiga}, {Marchal}, {Marrese}, {Moitinho}, {Mora}, {Muinonen}, {Osborne}, {Pancino}, {Pauwels}, {Petit}, {Recio-Blanco}, {Richards}, {Riello}, {Rimoldini}, {Robin}, {Roegiers}, {Rybizki}, {Sarro}, {Siopis}, {Smith}, {Sozzetti}, {Ulla}, {Utrilla}, {van Leeuwen}, {van Reeven}, {Abbas}, {Abreu Aramburu}, {Accart}, {Aerts}, {Aguado}, {Ajaj}, {Altavilla}, {{\'A}lvarez}, {{\'A}lvarez Cid-Fuentes}, {Alves}, {Anderson}, {Anglada Varela}, {Antoja}, {Audard}, {Baines}, {Baker}, {Balaguer-N{\'u}{\~n}ez}, {Balbinot}, {Balog}, {Barache}, {Barbato}, {Barros}, {Barstow}, {Bartolom{\'e}}, {Bassilana}, {Bauchet}, {Baudesson-Stella}, {Becciani}, {Bellazzini}, {Bernet}, {Bertone}, {Bianchi}, {Blanco-Cuaresma}, {Boch}, {Bombrun}, {Bossini},
  {Bouquillon}, {Bragaglia}, {Bramante}, {Breedt}, {Bressan}, {Brouillet}, {Bucciarelli}, {Burlacu}, {Busonero}, {Butkevich}, {Buzzi}, {Caffau}, {Cancelliere}, {C{\'a}novas}, {Cantat-Gaudin}, {Carballo}, {Carlucci}, {Carnerero}, {Carrasco}, {Casamiquela}, {Castellani}, {Castro-Ginard}, {Castro Sampol}, {Chaoul}, {Charlot}, {Chemin}, {Chiavassa}, {Cioni}, {Comoretto}, {Cooper}, {Cornez}, {Cowell}, {Crifo}, {Crosta}, {Crowley}, {Dafonte}, {Dapergolas}, {David}, {David}, {de Laverny}, {De Luise}, {De March}, {De Ridder}, {de Souza}, {de Teodoro}, {de Torres}, {del Peloso}, {del Pozo}, {Delbo}, {Delgado}, {Delgado}, {Delisle}, {Di Matteo}, {Diakite}, {Diener}, {Distefano}, {Dolding}, {Eappachen}, {Edvardsson}, {Enke}, {Esquej}, {Fabre}, {Fabrizio}, {Faigler}, {Fedorets}, {Fernique}, {Fienga}, {Figueras}, {Fouron}, {Fragkoudi}, {Fraile}, {Franke}, {Gai}, {Garabato}, {Garcia-Gutierrez}, {Garc{\'\i}a-Torres}, {Garofalo}, {Gavras}, {Gerlach}, {Geyer}, {Giacobbe}, {Gilmore}, {Girona}, {Giuffrida}, {Gomel}, {Gomez},
  {Gonzalez-Santamaria}, {Gonz{\'a}lez-Vidal}, {Granvik}, {Guti{\'e}rrez-S{\'a}nchez}, {Guy}, {Hauser}, {Haywood}, {Helmi}, {Hidalgo}, {Hilger}, {H{\l}adczuk}, {Hobbs}, {Holland}, {Huckle}, {Jasniewicz}, {Jonker}, {Juaristi Campillo}, {Julbe}, {Karbevska}, {Kervella}, {Khanna}, {Kochoska}, {Kontizas}, {Kordopatis}, {Korn}, {Kostrzewa-Rutkowska}, {Kruszy{\'n}ska}, {Lambert}, {Lanza}, {Lasne}, {Le Campion}, {Le Fustec}, {Lebreton}, {Lebzelter}, {Leccia}, {Leclerc}, {Lecoeur-Taibi}, {Liao}, {Licata}, {Lindstr{\o}m}, {Lister}, {Livanou}, {Lobel}, {Madrero Pardo}, {Managau}, {Mann}, {Marchant}, {Marconi}, {Marcos Santos}, {Marinoni}, {Marocco}, {Marshall}, {Martin Polo}, {Mart{\'\i}n-Fleitas}, {Masip}, {Massari}, {Mastrobuono-Battisti}, {Mazeh}, {McMillan}, {Messina}, {Michalik}, {Millar}, {Mints}, {Molina}, {Molinaro}, {Moln{\'a}r}, {Montegriffo}, {Mor}, {Morbidelli}, {Morel}, {Morris}, {Mulone}, {Munoz}, {Muraveva}, {Murphy}, {Musella}, {Noval}, {Ord{\'e}novic}, {Orr{\`u}}, {Osinde}, {Pagani}, {Pagano},
  {Palaversa}, {Palicio}, {Panahi}, {Pawlak}, {Pe{\~n}alosa Esteller}, {Penttil{\"a}}, {Piersimoni}, {Pineau}, {Plachy}, {Plum}, {Poggio}, {Poretti}, {Poujoulet}, {Pr{\v{s}}a}, {Pulone}, {Racero}, {Ragaini}, {Rainer}, {Raiteri}, {Rambaux}, {Ramos}, {Ramos-Lerate}, {Re Fiorentin}, {Regibo}, {Reyl{\'e}}, {Ripepi}, {Riva}, {Rixon}, {Robichon}, {Robin}, {Roelens}, {Rohrbasser}, {Romero-G{\'o}mez}, {Rowell}, {Royer}, {Rybicki}, {Sadowski}, {Sagrist{\`a} Sell{\'e}s}, {Sahlmann}, {Salgado}, {Salguero}, {Samaras}, {Sanchez Gimenez}, {Sanna}, {Santove{\~n}a}, {Sarasso}, {Schultheis}, {Sciacca}, {Segol}, {Segovia}, {S{\'e}gransan}, {Semeux}, {Shahaf}, {Siddiqui}, {Siebert}, {Siltala}, {Slezak}, {Smart}, {Solano}, {Solitro}, {Souami}, {Souchay}, {Spagna}, {Spoto}, {Steele}, {Steidelm{\"u}ller}, {Stephenson}, {S{\"u}veges}, {Szabados}, {Szegedi-Elek}, {Taris}, {Tauran}, {Taylor}, {Teixeira}, {Thuillot}, {Tonello}, {Torra}, {Torra}, {Turon}, {Unger}, {Vaillant}, {van Dillen}, {Vanel}, {Vecchiato}, {Viala}, {Vicente},
  {Voutsinas}, {Weiler}, {Wevers}, {Wyrzykowski}, {Yoldas}, {Yvard}, {Zhao}, {Zorec}, {Zucker}, {Zurbach}, \& {Zwitter}}]{gaiaedr3}
{Gaia Collaboration}, {Brown}, A.~G.~A., {Vallenari}, A., {et~al.} 2021, \aap, 649, A1, \dodoi{10.1051/0004-6361/202039657}

\bibitem[{{Gaia Collaboration} {et~al.}(2023){Gaia Collaboration}, {Vallenari}, {Brown}, {Prusti}, {de Bruijne}, {Arenou}, {Babusiaux}, {Biermann}, {Creevey}, {Ducourant}, {Evans}, {Eyer}, {Guerra}, {Hutton}, {Jordi}, {Klioner}, {Lammers}, {Lindegren}, {Luri}, {Mignard}, {Panem}, {Pourbaix}, {Randich}, {Sartoretti}, {Soubiran}, {Tanga}, {Walton}, {Bailer-Jones}, {Bastian}, {Drimmel}, {Jansen}, {Katz}, {Lattanzi}, {van Leeuwen}, {Bakker}, {Cacciari}, {Casta{\~n}eda}, {De Angeli}, {Fabricius}, {Fouesneau}, {Fr{\'e}mat}, {Galluccio}, {Guerrier}, {Heiter}, {Masana}, {Messineo}, {Mowlavi}, {Nicolas}, {Nienartowicz}, {Pailler}, {Panuzzo}, {Riclet}, {Roux}, {Seabroke}, {Sordo}, {Th{\'e}venin}, {Gracia-Abril}, {Portell}, {Teyssier}, {Altmann}, {Andrae}, {Audard}, {Bellas-Velidis}, {Benson}, {Berthier}, {Blomme}, {Burgess}, {Busonero}, {Busso}, {C{\'a}novas}, {Carry}, {Cellino}, {Cheek}, {Clementini}, {Damerdji}, {Davidson}, {de Teodoro}, {Nu{\~n}ez Campos}, {Delchambre}, {Dell'Oro}, {Esquej},
  {Fern{\'a}ndez-Hern{\'a}ndez}, {Fraile}, {Garabato}, {Garc{\'\i}a-Lario}, {Gosset}, {Haigron}, {Halbwachs}, {Hambly}, {Harrison}, {Hern{\'a}ndez}, {Hestroffer}, {Hodgkin}, {Holl}, {Jan{\ss}en}, {Jevardat de Fombelle}, {Jordan}, {Krone-Martins}, {Lanzafame}, {L{\"o}ffler}, {Marchal}, {Marrese}, {Moitinho}, {Muinonen}, {Osborne}, {Pancino}, {Pauwels}, {Recio-Blanco}, {Reyl{\'e}}, {Riello}, {Rimoldini}, {Roegiers}, {Rybizki}, {Sarro}, {Siopis}, {Smith}, {Sozzetti}, {Utrilla}, {van Leeuwen}, {Abbas}, {{\'A}brah{\'a}m}, {Abreu Aramburu}, {Aerts}, {Aguado}, {Ajaj}, {Aldea-Montero}, {Altavilla}, {{\'A}lvarez}, {Alves}, {Anders}, {Anderson}, {Anglada Varela}, {Antoja}, {Baines}, {Baker}, {Balaguer-N{\'u}{\~n}ez}, {Balbinot}, {Balog}, {Barache}, {Barbato}, {Barros}, {Barstow}, {Bartolom{\'e}}, {Bassilana}, {Bauchet}, {Becciani}, {Bellazzini}, {Berihuete}, {Bernet}, {Bertone}, {Bianchi}, {Binnenfeld}, {Blanco-Cuaresma}, {Blazere}, {Boch}, {Bombrun}, {Bossini}, {Bouquillon}, {Bragaglia}, {Bramante}, {Breedt},
  {Bressan}, {Brouillet}, {Brugaletta}, {Bucciarelli}, {Burlacu}, {Butkevich}, {Buzzi}, {Caffau}, {Cancelliere}, {Cantat-Gaudin}, {Carballo}, {Carlucci}, {Carnerero}, {Carrasco}, {Casamiquela}, {Castellani}, {Castro-Ginard}, {Chaoul}, {Charlot}, {Chemin}, {Chiaramida}, {Chiavassa}, {Chornay}, {Comoretto}, {Contursi}, {Cooper}, {Cornez}, {Cowell}, {Crifo}, {Cropper}, {Crosta}, {Crowley}, {Dafonte}, {Dapergolas}, {David}, {David}, {de Laverny}, {De Luise}, {De March}, {De Ridder}, {de Souza}, {de Torres}, {del Peloso}, {del Pozo}, {Delbo}, {Delgado}, {Delisle}, {Demouchy}, {Dharmawardena}, {Di Matteo}, {Diakite}, {Diener}, {Distefano}, {Dolding}, {Edvardsson}, {Enke}, {Fabre}, {Fabrizio}, {Faigler}, {Fedorets}, {Fernique}, {Fienga}, {Figueras}, {Fournier}, {Fouron}, {Fragkoudi}, {Gai}, {Garcia-Gutierrez}, {Garcia-Reinaldos}, {Garc{\'\i}a-Torres}, {Garofalo}, {Gavel}, {Gavras}, {Gerlach}, {Geyer}, {Giacobbe}, {Gilmore}, {Girona}, {Giuffrida}, {Gomel}, {Gomez}, {Gonz{\'a}lez-N{\'u}{\~n}ez},
  {Gonz{\'a}lez-Santamar{\'\i}a}, {Gonz{\'a}lez-Vidal}, {Granvik}, {Guillout}, {Guiraud}, {Guti{\'e}rrez-S{\'a}nchez}, {Guy}, {Hatzidimitriou}, {Hauser}, {Haywood}, {Helmer}, {Helmi}, {Sarmiento}, {Hidalgo}, {Hilger}, {H{\l}adczuk}, {Hobbs}, {Holland}, {Huckle}, {Jardine}, {Jasniewicz}, {Jean-Antoine Piccolo}, {Jim{\'e}nez-Arranz}, {Jorissen}, {Juaristi Campillo}, {Julbe}, {Karbevska}, {Kervella}, {Khanna}, {Kontizas}, {Kordopatis}, {Korn}, {K{\'o}sp{\'a}l}, {Kostrzewa-Rutkowska}, {Kruszy{\'n}ska}, {Kun}, {Laizeau}, {Lambert}, {Lanza}, {Lasne}, {Le Campion}, {Lebreton}, {Lebzelter}, {Leccia}, {Leclerc}, {Lecoeur-Taibi}, {Liao}, {Licata}, {Lindstr{\o}m}, {Lister}, {Livanou}, {Lobel}, {Lorca}, {Loup}, {Madrero Pardo}, {Magdaleno Romeo}, {Managau}, {Mann}, {Manteiga}, {Marchant}, {Marconi}, {Marcos}, {Marcos Santos}, {Mar{\'\i}n Pina}, {Marinoni}, {Marocco}, {Marshall}, {Martin Polo}, {Mart{\'\i}n-Fleitas}, {Marton}, {Mary}, {Masip}, {Massari}, {Mastrobuono-Battisti}, {Mazeh}, {McMillan}, {Messina}, {Michalik},
  {Millar}, {Mints}, {Molina}, {Molinaro}, {Moln{\'a}r}, {Monari}, {Mongui{\'o}}, {Montegriffo}, {Montero}, {Mor}, {Mora}, {Morbidelli}, {Morel}, {Morris}, {Muraveva}, {Murphy}, {Musella}, {Nagy}, {Noval}, {Oca{\~n}a}, {Ogden}, {Ordenovic}, {Osinde}, {Pagani}, {Pagano}, {Palaversa}, {Palicio}, {Pallas-Quintela}, {Panahi}, {Payne-Wardenaar}, {Pe{\~n}alosa Esteller}, {Penttil{\"a}}, {Pichon}, {Piersimoni}, {Pineau}, {Plachy}, {Plum}, {Poggio}, {Pr{\v{s}}a}, {Pulone}, {Racero}, {Ragaini}, {Rainer}, {Raiteri}, {Rambaux}, {Ramos}, {Ramos-Lerate}, {Re Fiorentin}, {Regibo}, {Richards}, {Rios Diaz}, {Ripepi}, {Riva}, {Rix}, {Rixon}, {Robichon}, {Robin}, {Robin}, {Roelens}, {Rogues}, {Rohrbasser}, {Romero-G{\'o}mez}, {Rowell}, {Royer}, {Ruz Mieres}, {Rybicki}, {Sadowski}, {S{\'a}ez N{\'u}{\~n}ez}, {Sagrist{\`a} Sell{\'e}s}, {Sahlmann}, {Salguero}, {Samaras}, {Sanchez Gimenez}, {Sanna}, {Santove{\~n}a}, {Sarasso}, {Schultheis}, {Sciacca}, {Segol}, {Segovia}, {S{\'e}gransan}, {Semeux}, {Shahaf}, {Siddiqui}, {Siebert},
  {Siltala}, {Silvelo}, {Slezak}, {Slezak}, {Smart}, {Snaith}, {Solano}, {Solitro}, {Souami}, {Souchay}, {Spagna}, {Spina}, {Spoto}, {Steele}, {Steidelm{\"u}ller}, {Stephenson}, {S{\"u}veges}, {Surdej}, {Szabados}, {Szegedi-Elek}, {Taris}, {Taylor}, {Teixeira}, {Tolomei}, {Tonello}, {Torra}, {Torra}, {Torralba Elipe}, {Trabucchi}, {Tsounis}, {Turon}, {Ulla}, {Unger}, {Vaillant}, {van Dillen}, {van Reeven}, {Vanel}, {Vecchiato}, {Viala}, {Vicente}, {Voutsinas}, {Weiler}, {Wevers}, {Wyrzykowski}, {Yoldas}, {Yvard}, {Zhao}, {Zorec}, {Zucker}, \& {Zwitter}}]{gaia2023}
{Gaia Collaboration}, {Vallenari}, A., {Brown}, A.~G.~A., {et~al.} 2023, \aap, 674, A1, \dodoi{10.1051/0004-6361/202243940}

\bibitem[{{Gaudi} {et~al.}(2005){Gaudi}, {Seager}, \& {Mallen-Ornelas}}]{2005ApJ...623..472G}
{Gaudi}, B.~S., {Seager}, S., \& {Mallen-Ornelas}, G. 2005, \apj, 623, 472, \dodoi{10.1086/428478}

\bibitem[{{Giacalone} {et~al.}(2017){Giacalone}, {Matsakos}, \& {K{\"o}nigl}}]{2017AJ....154..192G}
{Giacalone}, S., {Matsakos}, T., \& {K{\"o}nigl}, A. 2017, \aj, 154, 192, \dodoi{10.3847/1538-3881/aa8c04}

\bibitem[{{Goldreich} \& {Tremaine}(1979)}]{1979ApJ...233..857G}
{Goldreich}, P., \& {Tremaine}, S. 1979, \apj, 233, 857, \dodoi{10.1086/157448}

\bibitem[{{Gonz{\'a}lez Hern{\'a}ndez} {et~al.}(2024){Gonz{\'a}lez Hern{\'a}ndez}, {Su{\'a}rez Mascare{\~n}o}, {Silva}, {Stefanov}, {Faria}, {Tabernero}, {Sozzetti}, {Rebolo}, {Pepe}, {Santos}, {Cristiani}, {Lovis}, {Dumusque}, {Figueira}, {Lillo-Box}, {Nari}, {Benatti}, {Hobson}, {Castro-Gonz{\'a}lez}, {Allart}, {Passegger}, {Zapatero Osorio}, {Adibekyan}, {Alibert}, {Allende Prieto}, {Bouchy}, {Damasso}, {D'Odorico}, {Di Marcantonio}, {Ehrenreich}, {Lo Curto}, {Santos}, {Martins}, {Mehner}, {Micela}, {Molaro}, {Nunes}, {Palle}, {Sousa}, \& {Udry}}]{2024A&A...690A..79G}
{Gonz{\'a}lez Hern{\'a}ndez}, J.~I., {Su{\'a}rez Mascare{\~n}o}, A., {Silva}, A.~M., {et~al.} 2024, \aap, 690, A79, \dodoi{10.1051/0004-6361/202451311}

\bibitem[{Hartman \& Lépine(2020)}]{superwide_2020}
Hartman, Z.~D., \& Lépine, S. 2020, The Astrophysical Journal Supplement Series, 247, 66, \dodoi{10.3847/1538-4365/ab79a6}

\bibitem[{Helled {et~al.}(2014)Helled, Bodenheimer, Podolak, Boley, Meru, Nayakshin, Fortney, Mayer, Alibert, \& Boss}]{helledGiantPlanetFormation2014}
Helled, R., Bodenheimer, P., Podolak, M., {et~al.} 2014, Giant {{Planet Formation}}, {{Evolution}}, and {{Internal Structure}} (eprint: arXiv:1311.1142), 643--665, \dodoi{10.2458/azu_uapress_9780816531240-ch028}

\bibitem[{Huang {et~al.}(2020{\natexlab{a}})Huang, Vanderburg, Pál, Sha, Yu, Fong, Fausnaugh, Shporer, Guerrero, Vanderspek, \& Ricker}]{qlp1}
Huang, C.~X., Vanderburg, A., Pál, A., {et~al.} 2020{\natexlab{a}}, Research Notes of the AAS, 4, 204, \dodoi{10.3847/2515-5172/abca2e}

\bibitem[{Huang {et~al.}(2020{\natexlab{b}})Huang, Vanderburg, Pál, Sha, Yu, Fong, Fausnaugh, Shporer, Guerrero, Vanderspek, \& Ricker}]{qlp2}
---. 2020{\natexlab{b}}, Research Notes of the AAS, 4, 206, \dodoi{10.3847/2515-5172/abca2d}

\bibitem[{{Hu{\'e}lamo} {et~al.}(2008){Hu{\'e}lamo}, {Figueira}, {Bonfils}, {Santos}, {Pepe}, {Gillon}, {Azevedo}, {Barman}, {Fern{\'a}ndez}, {di Folco}, {Guenther}, {Lovis}, {Melo}, {Queloz}, \& {Udry}}]{2008A&A...489L...9H}
{Hu{\'e}lamo}, N., {Figueira}, P., {Bonfils}, X., {et~al.} 2008, \aap, 489, L9, \dodoi{10.1051/0004-6361:200810596}

\bibitem[{{Jeffers} {et~al.}(2018){Jeffers}, {Sch{\"o}fer}, {Lamert}, {Reiners}, {Montes}, {Caballero}, {Cort{\'e}s-Contreras}, {Marvin}, {Passegger}, {Zechmeister}, {Quirrenbach}, {Alonso-Floriano}, {Amado}, {Bauer}, {Casal}, {Diez Alonso}, {Herrero}, {Morales}, {Mundt}, {Ribas}, \& {Sarmiento}}]{2018A&A...614A..76J}
{Jeffers}, S.~V., {Sch{\"o}fer}, P., {Lamert}, A., {et~al.} 2018, \aap, 614, A76, \dodoi{10.1051/0004-6361/201629599}

\bibitem[{Jenkins {et~al.}(2016)Jenkins, Twicken, McCauliff, Campbell, Sanderfer, Lung, Mansouri-Samani, Girouard, Tenenbaum, Klaus, Smith, Caldwell, Chacon, Henze, Heiges, Latham, Morgan, Swade, Rinehart, \& Vanderspek}]{spoc}
Jenkins, J., Twicken, J., McCauliff, S., {et~al.} 2016, in Software and Cyber infrastructure for Astronomy IV, \dodoi{10.1117/12.2233418}

\bibitem[{Kempton {et~al.}(2018)Kempton, Bean, Louie, Deming, Koll, Mansfield, Christiansen, López-Morales, Swain, Zellem, Ballard, Barclay, Barstow, Batalha, Beatty, Berta-Thompson, Birkby, Buchhave, Charbonneau, Cowan, Crossfield, de~Val-Borro, Doyon, Dragomir, Gaidos, Heng, Hu, Kane, Kreidberg, Mallonn, Morley, Narita, Nascimbeni, Pallé, Quintana, Rauscher, Seager, Shkolnik, Sing, Sozzetti, Stassun, Valenti, \& von Essen}]{Kempton_2018}
Kempton, E. M.-R., Bean, J.~L., Louie, D.~R., {et~al.} 2018, Publications of the Astronomical Society of the Pacific, 130, 114401, \dodoi{10.1088/1538-3873/aadf6f}

\bibitem[{Kinoshita \& Nakai(2007)}]{kinoshitaGeneralSolutionKozai2007}
Kinoshita, H., \& Nakai, H. 2007, Celestial Mechanics and Dynamical Astronomy, 98, 67, \dodoi{10.1007/s10569-007-9069-6}

\bibitem[{Kozai(1962)}]{kozaiSecularPerturbationsAsteroids1962a}
Kozai, Y. 1962, The Astronomical Journal, 67, 591, \dodoi{10.1086/108790}

\bibitem[{{Kurucz}(1992)}]{kurucz1992}
{Kurucz}, R.~L. 1992, in The Stellar Populations of Galaxies, ed. B.~{Barbuy} \& A.~{Renzini}, Vol. 149, 225

\bibitem[{Lee \& Chiang(2015)}]{Lee_2015}
Lee, E.~J., \& Chiang, E. 2015, The Astrophysical Journal, 811, 41, \dodoi{10.1088/0004-637X/811/1/41}

\bibitem[{Lee \& Chiang(2016)}]{leeBreedingSuperEarthsBirthing2016}
---. 2016, The Astrophysical Journal, 817, 90, \dodoi{10.3847/0004-637X/817/2/90}

\bibitem[{{Leggett}(1992)}]{Leggett1992}
{Leggett}, S.~K. 1992, \apjs, 82, 351, \dodoi{10.1086/191720}

\bibitem[{Lidov(1962)}]{lidovEvolutionOrbitsArtificial1962}
Lidov, M.~L. 1962, Planetary and Space Science, 9, 719, \dodoi{10.1016/0032-0633(62)90129-0}

\bibitem[{{Lillo-Box} {et~al.}(2021){Lillo-Box}, {Faria}, {Su{\'a}rez Mascare{\~n}o}, {Figueira}, {Sousa}, {Tabernero}, {Lovis}, {Silva}, {Demangeon}, {Benatti}, {Santos}, {Mehner}, {Pepe}, {Sozzetti}, {Zapatero Osorio}, {Gonz{\'a}lez Hern{\'a}ndez}, {Micela}, {Hojjatpanah}, {Rebolo}, {Cristiani}, {Adibekyan}, {Allart}, {Allende Prieto}, {Cabral}, {Damasso}, {Di Marcantonio}, {Lo Curto}, {Martins}, {Megevand}, {Molaro}, {Nunes}, {Pall{\'e}}, {Pasquini}, {Poretti}, \& {Udry}}]{2021A&A...654A..60L}
{Lillo-Box}, J., {Faria}, J.~P., {Su{\'a}rez Mascare{\~n}o}, A., {et~al.} 2021, \aap, 654, A60, \dodoi{10.1051/0004-6361/202141714}

\bibitem[{{Lin} {et~al.}(1996){Lin}, {Bodenheimer}, \& {Richardson}}]{1996Natur.380..606L}
{Lin}, D.~N.~C., {Bodenheimer}, P., \& {Richardson}, D.~C. 1996, \nat, 380, 606, \dodoi{10.1038/380606a0}

\bibitem[{{Lindegren} {et~al.}(2021){Lindegren}, {Klioner, S. A.}, {Hernández, J.}, {Bombrun, A.}, {Ramos-Lerate, M.}, {Steidelmüller, H.}, {Bastian, U.}, {Biermann, M.}, {de Torres, A.}, {Gerlach, E.}, {Geyer, R.}, {Hilger, T.}, {Hobbs, D.}, {Lammers, U.}, {McMillan, P. J.}, {Stephenson, C. A.}, {Castañeda, J.}, {Davidson, M.}, {Fabricius, C.}, {Gracia-Abril, G.}, {Portell, J.}, {Rowell, N.}, {Teyssier, D.}, {Torra, F.}, {Bartolomé, S.}, {Clotet, M.}, {Garralda, N.}, {González-Vidal, J. J.}, {Torra, J.}, {Abbas, U.}, {Altmann, M.}, {Anglada Varela, E.}, {Balaguer-Núñez, L.}, {Balog, Z.}, {Barache, C.}, {Becciani, U.}, {Bernet, M.}, {Bertone, S.}, {Bianchi, L.}, {Bouquillon, S.}, {Brown, A. G. A.}, {Bucciarelli, B.}, {Busonero, D.}, {Butkevich, A. G.}, {Buzzi, R.}, {Cancelliere, R.}, {Carlucci, T.}, {Charlot, P.}, {Cioni, M.-R. L.}, {Crosta, M.}, {Crowley, C.}, {del Peloso, E. F.}, {del Pozo, E.}, {Drimmel, R.}, {Esquej, P.}, {Fienga, A.}, {Fraile, E.}, {Gai, M.}, {Garcia-Reinaldos, M.}, {Guerra, R.},
  {Hambly, N. C.}, {Hauser, M.}, {Janßen, K.}, {Jordan, S.}, {Kostrzewa-Rutkowska, Z.}, {Lattanzi, M. G.}, {Liao, S.}, {Licata, E.}, {Lister, T. A.}, {Löffler, W.}, {Marchant, J. M.}, {Masip, A.}, {Mignard, F.}, {Mints, A.}, {Molina, D.}, {Mora, A.}, {Morbidelli, R.}, {Murphy, C. P.}, {Pagani, C.}, {Panuzzo, P.}, {Peñalosa Esteller, X.}, {Poggio, E.}, {Re Fiorentin, P.}, {Riva, A.}, {Sagristà Sellés, A.}, {Sanchez Gimenez, V.}, {Sarasso, M.}, {Sciacca, E.}, {Siddiqui, H. I.}, {Smart, R. L.}, {Souami, D.}, {Spagna, A.}, {Steele, I. A.}, {Taris, F.}, {Utrilla, E.}, {van Reeven, W.}, \& {Vecchiato, A.}}]{Lindegren_2021}
{Lindegren}, {Klioner, S. A.}, {Hernández, J.}, {et~al.} 2021, A\&A, 649, A2, \dodoi{10.1051/0004-6361/202039709}

\bibitem[{Liu {et~al.}(2015)Liu, Mu{\~n}oz, \& Lai}]{liuSuppressionExtremeOrbital2015}
Liu, B., Mu{\~n}oz, D.~J., \& Lai, D. 2015, Monthly Notices of the Royal Astronomical Society, 447, 747, \dodoi{10.1093/mnras/stu2396}

\bibitem[{{Livingston} {et~al.}(2018){Livingston}, {Crossfield}, {Petigura}, {Gonzales}, {Ciardi}, {Beichman}, {Christiansen}, {Dressing}, {Henning}, {Howard}, {Isaacson}, {Fulton}, {Kosiarek}, {Schlieder}, {Sinukoff}, \& {Tamura}}]{2018AJ....156..277L}
{Livingston}, J.~H., {Crossfield}, I. J.~M., {Petigura}, E.~A., {et~al.} 2018, \aj, 156, 277, \dodoi{10.3847/1538-3881/aae778}

\bibitem[{Lopez \& Fortney(2014)}]{Lopez_2014}
Lopez, E.~D., \& Fortney, J.~J. 2014, The Astrophysical Journal, 792, 1, \dodoi{10.1088/0004-637X/792/1/1}

\bibitem[{Lundkvist {et~al.}(2016)Lundkvist, Kjeldsen, Albrecht, Davies, Basu, Huber, {Justesen}, {Karoff}, {Silva Aguirre}, {van Eylen}, {Vang}, {Arentoft}, {Barclay}, {Bedding}, {Campante}, {Chaplin}, {Christensen-Dalsgaard}, {Elsworth}, {Gilliland}, {Handberg}, {Hekker}, {Kawaler}, {Lund}, {Metcalfe}, {Miglio}, {Rowe}, {Stello}, {Tingley}, \& {White}}]{Lundkvist_16}
Lundkvist, M.~S., Kjeldsen, H., Albrecht, S., {et~al.} 2016, Nature Communications, 7, 11201, \dodoi{10.1038/ncomms11201}

\bibitem[{Madhusudhan {et~al.}(2014)Madhusudhan, Knutson, Fortney, \& Barman}]{Madhusudhan_2014}
Madhusudhan, N., Knutson, H., Fortney, J.~J., \& Barman, T. 2014, Exoplanetary Atmospheres (University of Arizona Press), \dodoi{10.2458/azu_uapress_9780816531240-ch032}

\bibitem[{{Mandel} \& {Agol}(2002)}]{Mandel2002}
{Mandel}, K., \& {Agol}, E. 2002, \apjl, 580, L171, \dodoi{10.1086/345520}

\bibitem[{{Matsakos} \& {K{\"o}nigl}(2016)}]{2016ApJ...820L...8M}
{Matsakos}, T., \& {K{\"o}nigl}, A. 2016, \apjl, 820, L8, \dodoi{10.3847/2041-8205/820/1/L8}

\bibitem[{{Mazeh} {et~al.}(2016){Mazeh}, {Holczer}, \& {Faigler}}]{Mazeh2016}
{Mazeh}, T., {Holczer}, T., \& {Faigler}, S. 2016, A\&A, 589, A75, \dodoi{10.1051/0004-6361/201528065}

\bibitem[{{McLaughlin}(1924)}]{McLaughlin_1924}
{McLaughlin}, D.~B. 1924, \apj, 60, 22, \dodoi{10.1086/142826}

\bibitem[{Mordasini {et~al.}(2015)Mordasini, Mollière, Dittkrist, Jin, \& Alibert}]{Mordasini_15}
Mordasini, C., Mollière, P., Dittkrist, K.-M., Jin, S., \& Alibert, Y. 2015, International Journal of Astrobiology, 14, 201–232, \dodoi{10.1017/S1473550414000263}

\bibitem[{Mordasini {et~al.}(2011)Mordasini, Mayor, Udry, Lovis, Ségransan, {Benz, W.}, {Bertaux, J.-L.}, {Bouchy, F.}, {Lo Curto, G.}, {Moutou, C.}, {Naef, D.}, {Pepe, F.}, {Queloz, D.}, \& {Santos, N. C.}}]{Mordasini_11}
Mordasini, C., Mayor, M., Udry, S., {et~al.} 2011, A\&A, 526, A111, \dodoi{10.1051/0004-6361/200913521}

\bibitem[{Naoz(2016)}]{2016ARA&A..54..441N}
Naoz, S. 2016, 54, 441, \dodoi{10.1146/annurev-astro-081915-023315}

\bibitem[{Naoz {et~al.}(2011)Naoz, Farr, Lithwick, Rasio, \& Teyssandier}]{2011Natur.473..187N}
Naoz, S., Farr, W.~M., Lithwick, Y., Rasio, F.~A., \& Teyssandier, J. 2011, 473, 187, \dodoi{10.1038/nature10076}

\bibitem[{{Otegi} {et~al.}(2020){Otegi}, {Bouchy}, \& {Helled}}]{Otegi20}
{Otegi}, J.~F., {Bouchy}, F., \& {Helled}, R. 2020, A\&A, 634, A43, \dodoi{10.1051/0004-6361/201936482}

\bibitem[{{Owen} \& {Lai}(2018)}]{Owen2018}
{Owen}, J.~E., \& {Lai}, D. 2018, \mnras, 479, 5012, \dodoi{10.1093/mnras/sty1760}

\bibitem[{Paegert {et~al.}(2021)Paegert, Stassun, Collins, Pepper, Torres, Jenkins, Twicken, \& Latham}]{ticv8.2}
Paegert, M., Stassun, K.~G., Collins, K.~A., {et~al.} 2021, TESS Input Catalog versions 8.1 and 8.2: Phantoms in the 8.0 Catalog and How to Handle Them.
\newblock \doarXiv{2108.04778}

\bibitem[{Petigura {et~al.}(2016)Petigura, Howard, Lopez, Deck, Fulton, Crossfield, Ciardi, Chiang, Lee, Isaacson, Beichman, Hansen, Schlieder, \& Sinukoff}]{Petigura_2016}
Petigura, E.~A., Howard, A.~W., Lopez, E.~D., {et~al.} 2016, The Astrophysical Journal, 818, 36, \dodoi{10.3847/0004-637X/818/1/36}

\bibitem[{Petigura {et~al.}(2017)Petigura, Sinukoff, Lopez, Crossfield, Howard, Brewer, Fulton, Isaacson, Ciardi, Howell, Everett, Horch, Hirsch, Weiss, \& Schlieder}]{Petigura_2017}
Petigura, E.~A., Sinukoff, E., Lopez, E.~D., {et~al.} 2017, The Astronomical Journal, 153, 142, \dodoi{10.3847/1538-3881/aa5ea5}

\bibitem[{Petrovich(2015)}]{petrovichHotJupitersCoplanar2015}
Petrovich, C. 2015, The Astrophysical Journal, 805, 75, \dodoi{10.1088/0004-637X/805/1/75}

\bibitem[{Pollack {et~al.}(1996)Pollack, Hubickyj, Bodenheimer, Lissauer, Podolak, \& Greenzweig}]{POLLACK_1996}
Pollack, J.~B., Hubickyj, O., Bodenheimer, P., {et~al.} 1996, Icarus, 124, 62, \dodoi{https://doi.org/10.1006/icar.1996.0190}

\bibitem[{Queloz {et~al.}(2001)Queloz, Henry, Sivan, Baliunas, Beuzit, {Donahue, R. A.}, {Mayor, M.}, {Naef, D.}, {Perrier, C.}, \& {Udry, S.}}]{Bisector1}
Queloz, D., Henry, G.~W., Sivan, J.~P., {et~al.} 2001, A\&A, 379, 279, \dodoi{10.1051/0004-6361:20011308}

\bibitem[{{Rossiter}(1924)}]{Rossiter_1924}
{Rossiter}, R.~A. 1924, \apj, 60, 15, \dodoi{10.1086/142825}

\bibitem[{{Santos} {et~al.}(2014){Santos}, {Mortier}, {Faria}, {Dumusque}, {Adibekyan}, {Delgado-Mena}, {Figueira}, {Benamati}, {Boisse}, {Cunha}, {Gomes da Silva}, {Lo Curto}, {Lovis}, {Martins}, {Mayor}, {Melo}, {Oshagh}, {Pepe}, {Queloz}, {Santerne}, {S{\'e}gransan}, {Sozzetti}, {Sousa}, \& {Udry}}]{2014A&A...566A..35S}
{Santos}, N.~C., {Mortier}, A., {Faria}, J.~P., {et~al.} 2014, \aap, 566, A35, \dodoi{10.1051/0004-6361/201423808}

\bibitem[{{Schlafly} \& {Finkbeiner}(2011)}]{extinction}
{Schlafly}, E.~F., \& {Finkbeiner}, D.~P. 2011, \apj, 737, 103, \dodoi{10.1088/0004-637X/737/2/103}

\bibitem[{{Sch{\"o}nrich} {et~al.}(2010){Sch{\"o}nrich}, {Binney}, \& {Dehnen}}]{Sch2010}
{Sch{\"o}nrich}, R., {Binney}, J., \& {Dehnen}, W. 2010, \mnras, 403, 1829, \dodoi{10.1111/j.1365-2966.2010.16253.x}

\bibitem[{{Smith} {et~al.}(2012){Smith}, {Stumpe}, {Van Cleve}, {Jenkins}, {Barclay}, {Fanelli}, {Girouard}, {Kolodziejczak}, {McCauliff}, {Morris}, \& {Twicken}}]{smith_2012}
{Smith}, J.~C., {Stumpe}, M.~C., {Van Cleve}, J.~E., {et~al.} 2012, \pasp, 124, 1000, \dodoi{10.1086/667697}

\bibitem[{Southworth(2011)}]{TEPcat}
Southworth, J. 2011, Monthly Notices of the Royal Astronomical Society, 417, 2166, \dodoi{10.1111/j.1365-2966.2011.19399.x}

\bibitem[{{Stassun} \& {Torres}(2016)}]{SED1}
{Stassun}, K.~G., \& {Torres}, G. 2016, \apjl, 831, L6, \dodoi{10.3847/2041-8205/831/1/L6}

\bibitem[{{Stassun} {et~al.}(2018){Stassun}, {Oelkers}, {Pepper}, {Paegert}, {De Lee}, {Torres}, {Latham}, {Charpinet}, {Dressing}, {Huber}, {Kane}, {L{\'e}pine}, {Mann}, {Muirhead}, {Rojas-Ayala}, {Silvotti}, {Fleming}, {Levine}, \& {Plavchan}}]{2018AJ....156..102S}
{Stassun}, K.~G., {Oelkers}, R.~J., {Pepper}, J., {et~al.} 2018, \aj, 156, 102, \dodoi{10.3847/1538-3881/aad050}

\bibitem[{Storch {et~al.}(2014)Storch, Anderson, \& Lai}]{storchChaoticDynamicsStellar2014}
Storch, N.~I., Anderson, K.~R., \& Lai, D. 2014, Science, 345, 1317, \dodoi{10.1126/science.1254358}

\bibitem[{Storch \& Lai(2015)}]{storchChaoticDynamicsStellar2015}
Storch, N.~I., \& Lai, D. 2015, Monthly Notices of the Royal Astronomical Society, 448, 1821, \dodoi{10.1093/mnras/stv119}

\bibitem[{{Sulis} {et~al.}(2023){Sulis}, {Lendl}, {Cegla}, {Rodr{\'\i}guez D{\'\i}az}, {Bigot}, {Van Grootel}, {Bekkelien}, {Cameron}, {Maxted}, {Simon}, {Lovis}, {Scandariato}, {Bruno}, {Nardiello}, {Bonfanti}, {Fridlund}, {Persson}, {Salmon}, {Sousa}, {Wilson}, {Krenn}, {Hoyer}, {Santerne}, {Ehrenreich}, {Alibert}, {Alonso}, {Anglada}, {B{\'a}rczy}, {Barrado y Navascues}, {Barros}, {Baumjohann}, {Beck}, {Beck}, {Benz}, {Billot}, {Bonfils}, {Borsato}, {Brandeker}, {Broeg}, {Cabrera}, {Charnoz}, {Corral van Damme}, {Csizmadia}, {Davies}, {Deleuil}, {Deline}, {Delrez}, {Demangeon}, {Demory}, {Erikson}, {Fortier}, {Fossati}, {Gandolfi}, {Gillon}, {G{\"u}del}, {Heng}, {Isaak}, {Kiss}, {Laskar}, {Lecavelier des Etangs}, {Magrin}, {Munari}, {Nascimbeni}, {Olofsson}, {Ottensamer}, {Pagano}, {Pall{\'e}}, {Peter}, {Piotto}, {Pollacco}, {Queloz}, {Ragazzoni}, {Rando}, {Rauer}, {Ribas}, {Rieder}, {Santos}, {S{\'e}gransan}, {Smith}, {Steinberger}, {Steller}, {Szab{\'o}}, {Thomas}, {Udry}, {Walton}, \&
  {Wolter}}]{2023A&A...670A..24S}
{Sulis}, S., {Lendl}, M., {Cegla}, H.~M., {et~al.} 2023, \aap, 670, A24, \dodoi{10.1051/0004-6361/202244223}

\bibitem[{{Szab{\'o}} \& {Kiss}(2011)}]{2011ApJ...727L..44S}
{Szab{\'o}}, G.~M., \& {Kiss}, L.~L. 2011, \apjl, 727, L44, \dodoi{10.1088/2041-8205/727/2/L44}

\bibitem[{Teyssandier {et~al.}(2013)Teyssandier, Naoz, Lizarraga, \& Rasio}]{2013ApJ...779..166T}
Teyssandier, J., Naoz, S., Lizarraga, I., \& Rasio, F.~A. 2013, 779, 166, \dodoi{10.1088/0004-637X/779/2/166}

\bibitem[{{Thompson} {et~al.}(2018){Thompson}, {Coughlin}, {Hoffman}, {Mullally}, {Christiansen}, {Burke}, {Bryson}, {Batalha}, {Haas}, {Catanzarite}, {Rowe}, {Barentsen}, {Caldwell}, {Clarke}, {Jenkins}, {Li}, {Latham}, {Lissauer}, {Mathur}, {Morris}, {Seader}, {Smith}, {Klaus}, {Twicken}, {Van Cleve}, {Wohler}, {Akeson}, {Ciardi}, {Cochran}, {Henze}, {Howell}, {Huber}, {Pr{\v{s}}a}, {Ram{\'\i}rez}, {Morton}, {Barclay}, {Campbell}, {Chaplin}, {Charbonneau}, {Christensen-Dalsgaard}, {Dotson}, {Doyle}, {Dunham}, {Dupree}, {Ford}, {Geary}, {Girouard}, {Isaacson}, {Kjeldsen}, {Quintana}, {Ragozzine}, {Shabram}, {Shporer}, {Silva Aguirre}, {Steffen}, {Still}, {Tenenbaum}, {Welsh}, {Wolfgang}, {Zamudio}, {Koch}, \& {Borucki}}]{2018ApJS..235...38T}
{Thompson}, S.~E., {Coughlin}, J.~L., {Hoffman}, K., {et~al.} 2018, \apjs, 235, 38, \dodoi{10.3847/1538-4365/aab4f9}

\bibitem[{{Tokovinin}(2018)}]{2018tokovinin}
{Tokovinin}, A. 2018, \pasp, 130, 035002, \dodoi{10.1088/1538-3873/aaa7d9}

\bibitem[{{Torres} {et~al.}(2008){Torres}, {Winn}, \& {Holman}}]{Torres2008}
{Torres}, G., {Winn}, J.~N., \& {Holman}, M.~J. 2008, \apj, 677, 1324, \dodoi{10.1086/529429}

\bibitem[{{Triaud}(2018)}]{2018haex.bookE...2T}
{Triaud}, A. H.~M.~J. 2018, in Handbook of Exoplanets, ed. H.~J. {Deeg} \& J.~A. {Belmonte}, 2, \dodoi{10.1007/978-3-319-55333-7_2}

\bibitem[{{Udry} {et~al.}(2003){Udry}, {Mayor}, \& {Santos}}]{2003A&A...407..369U}
{Udry}, S., {Mayor}, M., \& {Santos}, N.~C. 2003, \aap, 407, 369, \dodoi{10.1051/0004-6361:20030843}

\bibitem[{{Vissapragada} {et~al.}(2022){Vissapragada}, {Knutson}, {Greklek-McKeon}, {Oklop{\v{c}}i{\'c}}, {Dai}, {dos Santos}, {Jovanovic}, {Mawet}, {Millar-Blanchaer}, {Paragas}, {Spake}, {Tinyanont}, \& {Vasisht}}]{2022AJ....164..234V}
{Vissapragada}, S., {Knutson}, H.~A., {Greklek-McKeon}, M., {et~al.} 2022, \aj, 164, 234, \dodoi{10.3847/1538-3881/ac92f2}

\bibitem[{von Zeipel(1910)}]{vonzeipelLapplicationSeriesLindstedt1910}
von Zeipel, H. 1910, \dodoi{10.1002/asna.19091832202}

\bibitem[{Wu \& Lithwick(2011)}]{2011ApJ...735..109W}
Wu, Y., \& Lithwick, Y. 2011, 735, 109, \dodoi{10.1088/0004-637X/735/2/109}

\bibitem[{Wu \& Murray(2003)}]{wuPlanetMigrationBinary2003}
Wu, Y., \& Murray, N. 2003, The Astrophysical Journal, 589, 605, \dodoi{10.1086/374598}

\bibitem[{Wu {et~al.}(2007)Wu, Murray, \& Ramsahai}]{2007ApJ...670..820W}
Wu, Y., Murray, N.~W., \& Ramsahai, J.~M. 2007, 670, 820, \dodoi{10.1086/521996}

\bibitem[{{Zechmeister} \& {K{\"u}rster}(2009)}]{periodogram}
{Zechmeister}, M., \& {K{\"u}rster}, M. 2009, \aap, 496, 577, \dodoi{10.1051/0004-6361:200811296}

\bibitem[{Zeng {et~al.}(2021)Zeng, Jacobsen, Hyung, Levi, Nava, Kirk, Piaulet, Lacedelli, Sasselov, Petaev, Stewart, Alam, López-Morales, Damasso, \& Latham}]{Zeng_2021}
Zeng, L., Jacobsen, S.~B., Hyung, E., {et~al.} 2021, The Astrophysical Journal, 923, 247, \dodoi{10.3847/1538-4357/ac3137}

\bibitem[{{Ziegler} {et~al.}(2020){Ziegler}, {Tokovinin}, {Brice{\~n}o}, {Mang}, {Law}, \& {Mann}}]{2020ziegler}
{Ziegler}, C., {Tokovinin}, A., {Brice{\~n}o}, C., {et~al.} 2020, \aj, 159, 19, \dodoi{10.3847/1538-3881/ab55e9}

\bibitem[{{Šubjak} {et~al.}(2022){Šubjak}, {Endl, M.}, {Chaturvedi, P.}, {Karjalainen, R.}, {Cochran, W. D.}, {Esposito, M.}, {Gandolfi, D.}, {Lam, K. W. F.}, {Stassun, K.}, {Žák, J.}, {Lodieu, N.}, {Boffin, H. M. J.}, {MacQueen, P. J.}, {Hatzes, A.}, {Guenther, E. W.}, {Georgieva, I.}, {Grziwa, S.}, {Schmerling, H.}, {Skarka, M.}, {Blažek, M.}, {Karjalainen, M.}, {Špoková, M.}, {Isaacson, H.}, {Howard, A. W.}, {Burke, C. J.}, {Van Eylen, V.}, {Falk, B.}, {Fridlund, M.}, {Goffo, E.}, {Jenkins, J. M.}, {Korth, J.}, {Lissauer, J. J.}, {Livingston, J. H.}, {Luque, R.}, {Muresan, A.}, {Osborn, H. P.}, {Pallé, E.}, {Persson, C. M.}, {Redfield, S.}, {Ricker, G. R.}, {Seager, S.}, {Serrano, L. M.}, {Smith, A. M. S.}, \& {Kabáth, P.}}]{subjak2022}
{Šubjak}, J., {Endl, M.}, {Chaturvedi, P.}, {et~al.} 2022, A\&A, 662, A107, \dodoi{10.1051/0004-6361/202142883}

\end{thebibliography}
\bibliographystyle{aasjournal}


\appendix


\section{Additional table}

\begin{table*}[h]
\centering
\caption{Radial velocity measurements of TOI-6038~A with PARAS-2{\textdagger}.}
\label{tab:rv_table}
\begin{tabular}{ccccc}
\hline
\hline
\noalign{\smallskip}
BJD$_{TDB}$& Relative-RV & $\sigma$-RV & BIS & $\sigma$-BIS\\
Days & m s$^{-1}$ & m s$^{-1}$ & m s$^{-1}$ & m s$^{-1}$\\
\noalign{\smallskip}
\hline
\noalign{\smallskip}
2459976.184815 & 0.50 & 21.66 & -277.05 & 26.26 \\
2459977.163062 & -29.79 & 11.46 & -118 & 17.82 \\
2459978.168539 & -10.76 & 15.59 & -141.83 & 21.75 \\
2459979.168275 & 8.55 & 15.77 & -0.05 & 30.58 \\
2460265.258637 & 11.36 & 6.45 & -56.86 & 3.46 \\
2460266.319432 & 21.73 & 8.26 & -105.16 & 12.98 \\
2460269.355211 & -13.86 & 9.92 & -153.42 & 9.42 \\
2460271.357469 & 13.31 & 9.33 & -67.79 & 10.88 \\
2460272.341773 & 2.53 & 9.36 & -82.45 & 26.69 \\
2460273.301944 & 17.08 & 6.93 & -38.28 & 14.33 \\
2460307.269484 & 12.04 & 8.77 & -34.44 & 5.91 \\
2460308.209118 & 3.63 & 4.62 & -7.38 & 5.95 \\
2460308.259393 & 14.69 & 6.33 & -10.88 & 7.52 \\
2460309.217266 & -32.06 & 6.63 & -10.75 & 5.46 \\
2460309.318800 & -22.49 & 10.43 & -17.67 & 19.71 \\
2460315.212541 & -20.22 & 8.68 & 39.02 & 8.05 \\
2460317.174921 & 2.13 & 9.52 & -69.26 & 15.6 \\
2460319.260228 & 24.94 & 7.23 & -148.22 & 10.46 \\
2460321.106960 & -14.63 & 4.77 & -168.22 & 11.3 \\
2460322.164151 & -26.85 & 5.04 & -126.97 & 6.37 \\
2460324.170865 & 35.07 & 5.72 & -174.63 & 7.45 \\
2460325.186100 & 17.49 & 5.99 & -142.68 & 7.75 \\
2460326.182065 & -6.42 & 11.37 & -198.02 & 16.09 \\
2460326.229318 & -1.97 & 17.29 & -262.08 & 26.84 \\
2460329.198876 & 3.88 & 11.54 & -343.74 & 18.87 \\
2460330.243780 & 21.17 & 8.07 & -171.41 & 9.79 \\
2460331.146195 & 2.67 & 6.99 & -248.43 & 9.06 \\
2460332.200807 & -2.94 & 6.08 & -197.37 & 12.51 \\
2460334.134084 & -21.00 & 6.71 & -176.11 & 8.61 \\
\noalign{\smallskip}
\hline
\noalign{\smallskip}
\multicolumn{5}{l}{\footnotesize{\textbf{Notes.} {\textdagger}Exposure time for each spectra is 3600 seconds.}}
\end{tabular}
\end{table*}


\newpage

\renewcommand{\arraystretch}{1.3}
\begin{table*}[t!]
\caption{Summary of EXOFASTv2 fitted and derived parameters for the TOI-6038~A system, based on the low-mass solution (relative probability = 13\%).}
\label{tab:exofast_table_lowmass_solution}   
\centering
\begin{tabular}{llll}
\hline
\hline
\noalign{\smallskip}
Parameter&Description&Value\\
\noalign{\smallskip}
\hline
\noalign{\smallskip}
\multicolumn{2}{l}{Stellar Parameters:}\\
~~~~$M_*$\dotfill &Mass (\msun)\dotfill &$1.145^{+0.015}_{-0.020}$\\
~~~~$R_*$\dotfill &Radius (\rsun)\dotfill &$1.623^{+0.059}_{-0.051}$\\
~~~~$L_*$\dotfill &Luminosity (\lsun)\dotfill &$3.24^{+0.26}_{-0.21}$\\
~~~~$\rho_*$\dotfill &Density (cgs)\dotfill &$0.376^{+0.037}_{-0.038}$\\
~~~~$T_{\rm eff}$\dotfill &Effective temperature (K)\dotfill &$6076^{+93}_{-95}$\\
~~~~$\log{g}$\dotfill &Surface gravity (cgs)\dotfill &$4.075^{+0.027}_{-0.030}$\\
~~~~$[{\rm Fe/H}]$\dotfill &Metallicity (dex)\dotfill &$0.077^{+0.069}_{-0.064}$\\
~~~~$Age$\dotfill &Age (Gyr)\dotfill &$6.31^{+0.43}_{-0.48}$\\
\smallskip\\\multicolumn{2}{l}{Planetary Parameters:}& \smallskip\\
~~~~$P$\dotfill &Period (days)\dotfill &$5.8267313^{+0.0000073}_{-0.0000067}$\\
~~~~$R_P$\dotfill &Radius ($R_\oplus$)\dotfill &$6.34^{+0.27}_{-0.24}$\\
~~~~$M_P$\dotfill &Mass ($M_\oplus$)\dotfill &$72.5^{+8.3}_{-9.2}$\\
~~~~$T_C$\dotfill &Observed Time of conjunction (\bjdtdb)\dotfill &$2459883.05804^{+0.00074}_{-0.00075}$\\
~~~~$a$\dotfill &Semi-major axis (AU)\dotfill &$0.06630^{+0.00029}_{-0.00039}$\\
~~~~$i$\dotfill &Inclination (Degrees)\dotfill &$88.10^{+0.97}_{-0.69}$\\
~~~~$T_{\rm eq}$\dotfill &Equilibrium temp{\textdagger} (K)\dotfill &$1450^{+28}_{-23}$\\
~~~~$R_P/R_*$\dotfill &Radius of planet in stellar radii \dotfill &$0.03579^{+0.00040}_{-0.00036}$\\
~~~~$a/R_*$\dotfill &Semi-major axis in stellar radii \dotfill &$8.77^{+0.28}_{-0.30}$\\
~~~~$\delta$\dotfill &$\left(R_P/R_*\right)^2$ \dotfill &$0.001281^{+0.000029}_{-0.000026}$\\
~~~~$b$\dotfill &Transit impact parameter \dotfill &$0.291^{+0.092}_{-0.14}$\\
~~~~$\rho_P$\dotfill &Density (cgs)\dotfill &$1.55^{+0.25}_{-0.24}$\\
~~~~$logg_P$\dotfill &Surface gravity (cgs)\dotfill &$3.244^{+0.057}_{-0.065}$\\
~~~~$\fave$\dotfill &Incident Flux (\fluxcgs)\dotfill &$1.004^{+0.079}_{-0.063}$\\
~~~~$M_P\sin i$\dotfill &Minimum mass ($M_\oplus$)\dotfill &$72.5^{+8.3}_{-9.2}$\\
~~~~$M_P/M_*$\dotfill &Mass ratio \dotfill &$0.000191^{+0.000022}_{-0.000024}$\\
~~~~$d/R_*$\dotfill &Separation at mid transit \dotfill &$8.77^{+0.28}_{-0.30}$\\
\noalign{\smallskip}
\hline
\noalign{\smallskip}
\multicolumn{3}{l}{\footnotesize{\textbf{Notes.} {\textdagger}Assumes no albedo and perfect redistribution}}
\end{tabular}
\end{table*}


\newpage

\section{Additional figure}

\begin{figure*}[h!]
\centering
\caption{Corner plot showing the covariances among the fitted parameters from EXOFASTv2 for the high-mass solution. The inner and outer contours around the median denote the 68\% and 95\% confidence intervals, respectively.}
\label{fig:corner_plot}
\includegraphics[width=\textwidth]{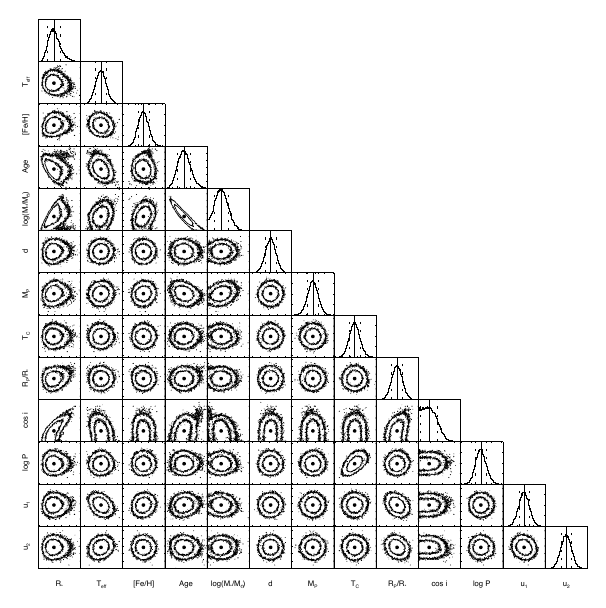}
\end{figure*}


\end{document}